\documentclass{vgtc}                          
\pdfoutput=1

\ifpdf
  \pdfoutput=1\relax                   
  \usepackage{graphicx}                
  \DeclareGraphicsExtensions{.pdf,.png,.jpg,.jpeg} 
\else
  \usepackage{graphicx}                
  \DeclareGraphicsExtensions{.eps}     
\fi%

\graphicspath{{./}} 

\usepackage{microtype}                 
\PassOptionsToPackage{warn}{textcomp}  
\usepackage{times}                     
\usepackage{cite}                      
\usepackage{tabu}                      
\usepackage{booktabs}                  

\usepackage{makecell}
\usepackage{amsmath}
\usepackage{overpic}

\usepackage{pifont}
\usepackage{authblk}
\usepackage{amssymb}
\usepackage{mathtools}
\usepackage{siunitx}
\usepackage{algorithmic}
\usepackage{algorithm}
\usepackage{listings}
\usepackage{multirow}
\usepackage{boldline}
\usepackage{array}
\newcolumntype{C}[1]{>{\centering\arraybackslash}p{#1}}
\usepackage{caption}
\usepackage{url}
\usepackage{ragged2e}
\usepackage{float}
\usepackage{subfig}
\onlineid{0}

\vgtccategory{Research}

\vgtcinsertpkg

\title{Realistic Volume Rendering with Environment-Synced \\ Illumination in Mixed Reality \vspace{-0.20cm}}

\author[1]{Haojie Cheng$^{*}$\thanks{$^{*}$Equal contribution}}
\author[1]{Chunxiao Xu$^{*}$\thanks{$^{\dag}$Corresponding author: hitic@sibet.ac.cn}}
\author[1]{Xujing Chen}
\author[1,2]{Zhenxin Chen}
\author[1,2]{Jiajun Wang}
\author[1,2]{Lingxiao Zhao$^{\dag}$}
\affil[1]{School of Biomedical Engineering, University of Science and Technology of China}
\affil[2]{Suzhou Institute of Biomedical Engineering and Technology, Chinese Academy of Sciences \vspace{-0.5cm}}

\makeatletter
\def\thanks#1{\protected@xdef\@thanks{\@thanks
        \protect\footnotetext{#1}}}
\makeatother

\teaser{
  \centering
  \begin{minipage}[b]{0.29\textwidth}
  \centering
  \includegraphics[width=\textwidth]{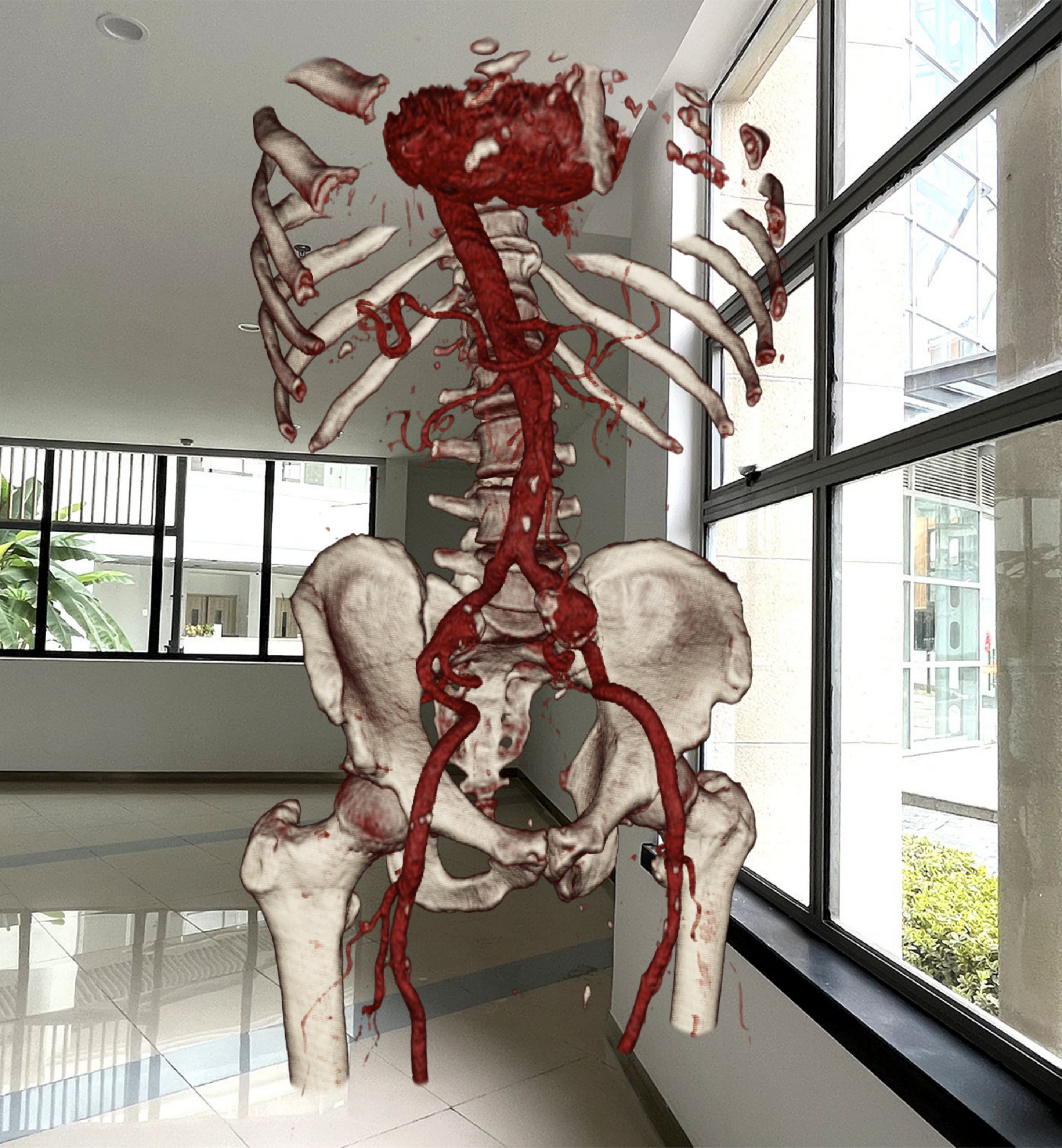}
  \end{minipage}
\begin{minipage}[b]{0.29\textwidth}
  \centering
  \includegraphics[width=\textwidth]{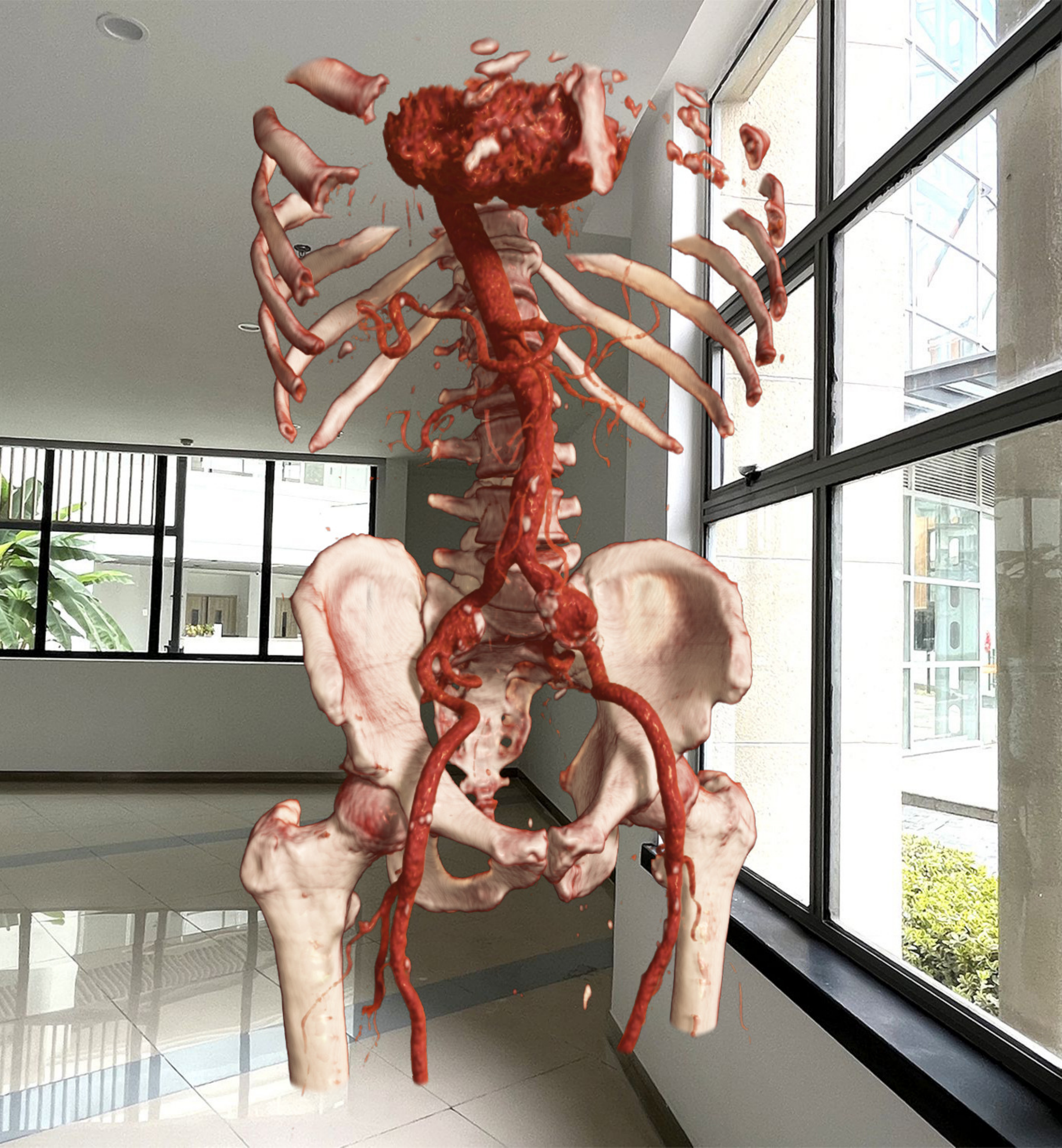}
  \end{minipage}
  \begin{minipage}[b]{0.3314\textwidth}
  \centering
  \includegraphics[width=\textwidth]{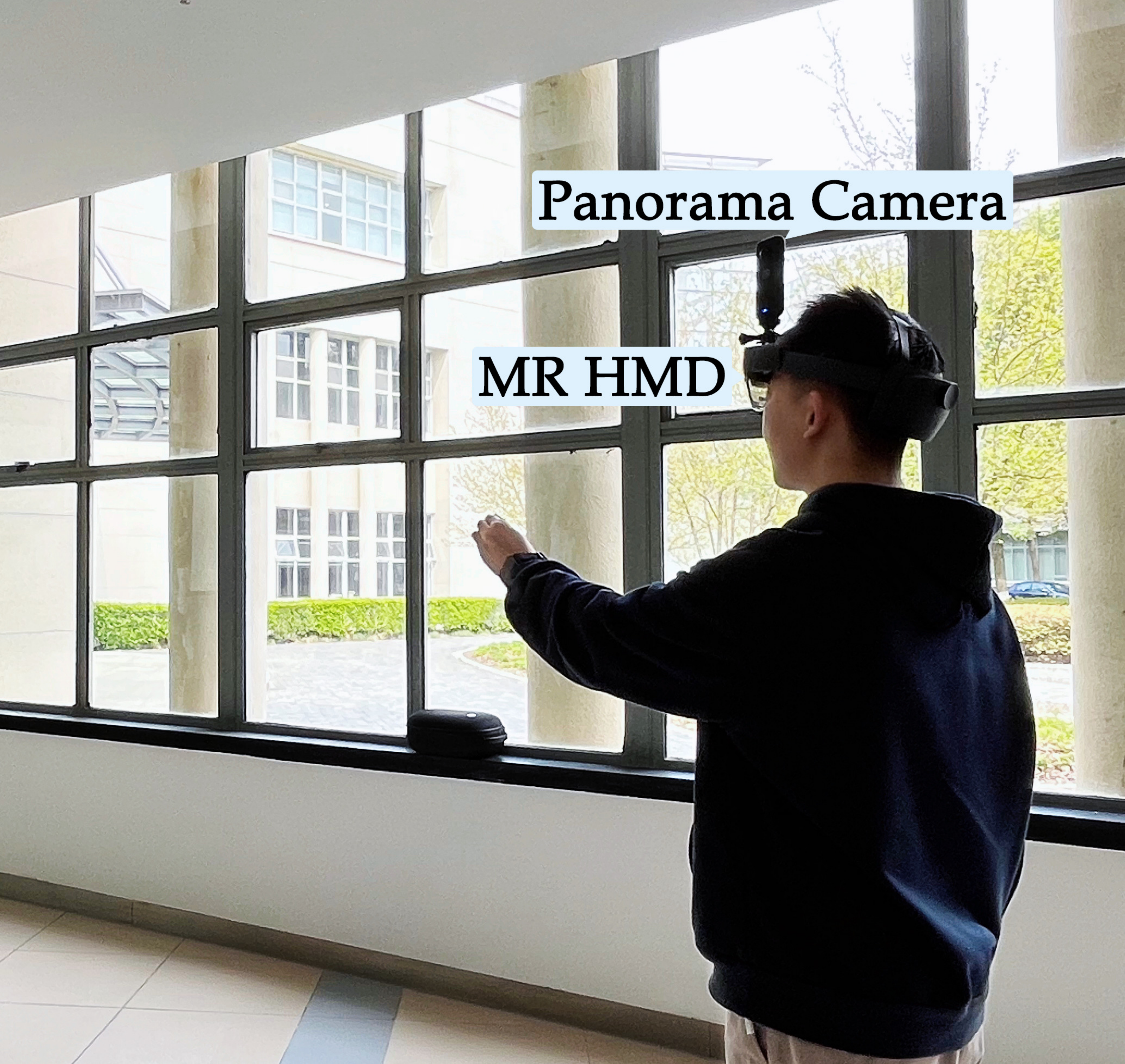}
  \end{minipage}
\caption{Visual comparison of MR volume rendering without (left) and with (middle) environment-synced illumination. \label{fig:teaser}
}}

\abstract{Interactive volume visualization using a mixed reality (MR) system helps provide users with an intuitive spatial perception of volumetric data. Due to sophisticated requirements of user interaction and vision when using MR head-mounted display (HMD) devices, the conflict between the realisticness and efficiency of direct volume rendering (DVR) is yet to be resolved. In this paper, a new MR visualization framework that supports interactive realistic DVR is proposed. An efficient illumination estimation method is used to identify the high dynamic range (HDR) environment illumination captured using a
panorama camera. To improve the visual quality of Monte Carlo-based DVR, a new spatio-temporal denoising algorithm is designed. Based on a reprojection strategy, it makes full use of temporal coherence between adjacent frames and spatial coherence between the two screens of an HMD to optimize MR rendering quality. Several MR development modules are also developed for related devices to efficiently and stably display the DVR results in an MR HMD. Experimental results demonstrate that our framework can better support immersive and intuitive user perception during MR viewing than existing MR solutions.} 

\CCScatlist{
\CCScatTwelve{Computing methodologies}{Computer graphics}{Graphics systems and interfaces}{Mixed / augmented reality};
\CCScatTwelve{Computing methodologies}{Computer graphics}{Rendering}{}
}

\begin{document}

\maketitle

\vspace{0.10in}
\section{Introduction}
\label{sec:intro}

Mixed reality (MR) interfaces including head-mounted display (HMD) devices are increasingly popular. Environment illumination is a crucial factor for improving the user's spatial perception of $3$D virtual models in MR \cite{Rhee17, Chalmers2020}. When the environment scene around virtual models varies over time, the shading effects of virtual models should synchronously reflect the surrounding illumination in real-time. In such a way, the user's sense of immersion during MR viewing can be significantly improved (Fig. \ref{fig:teaser}).

In scientific visualization, volumetric data plays an important role and is commonly used to retain raw information of scientific measurements or calibrations, such as medical images \cite{Elshafei2019}. Direct volume rendering (DVR) is commonly used for reconstructing and displaying of $3$D structures embedded in volumetric data \cite{Preim2013visual}. Realistic DVR simulates light transportation within volume space including absorption and scattering and is usually solved by the Monte Carlo (MC) based volumetric path tracing (VPT) strategy \cite{Kroes2012interactive, Von2016Efficient}. However, obtaining high-quality DVR results using these methods greatly relies on sufficiently sampling along a large number of light paths, and is normally very time-consuming. Due to the requirement of high interaction efficiency, most existing real-time DVR methods for VR, AR or MR applications \cite{Li20213d,Scholl2018direct, Waschk2020favr} preferred to make use of a non-physically ray-casting method without advanced volume illumination strategies. 

Previous research works \cite{Zhang2013,Von2016Efficient, jonsson2014survey} have shown that environment illumination is crucial for DVR and facilitates generating natural shading effects (e.g., soft shadow, ambient occlusion, etc.). It greatly enhances the surface detail of volumetric data with complex $3$D structures \cite{Diaz2017Experimental, lindemann2011influence}. Siemens researchers further demonstrated that the realistic DVR technique is valuable for improving the understanding of complex anatomical situations \cite{Elshafei2019}. However, their methods need a few seconds to obtain low-noise realistic DVR results and are only suitable for offline rendering tasks. Moreover, existing commercial rendering engines that support MR development (e.g., Unity$3$D or UE) also do not support real-time realistic DVR using their built-in rendering algorithms.

\begin{figure*}[t]
\centering
\hspace{-0.05in}
\includegraphics[width=1.0\textwidth]{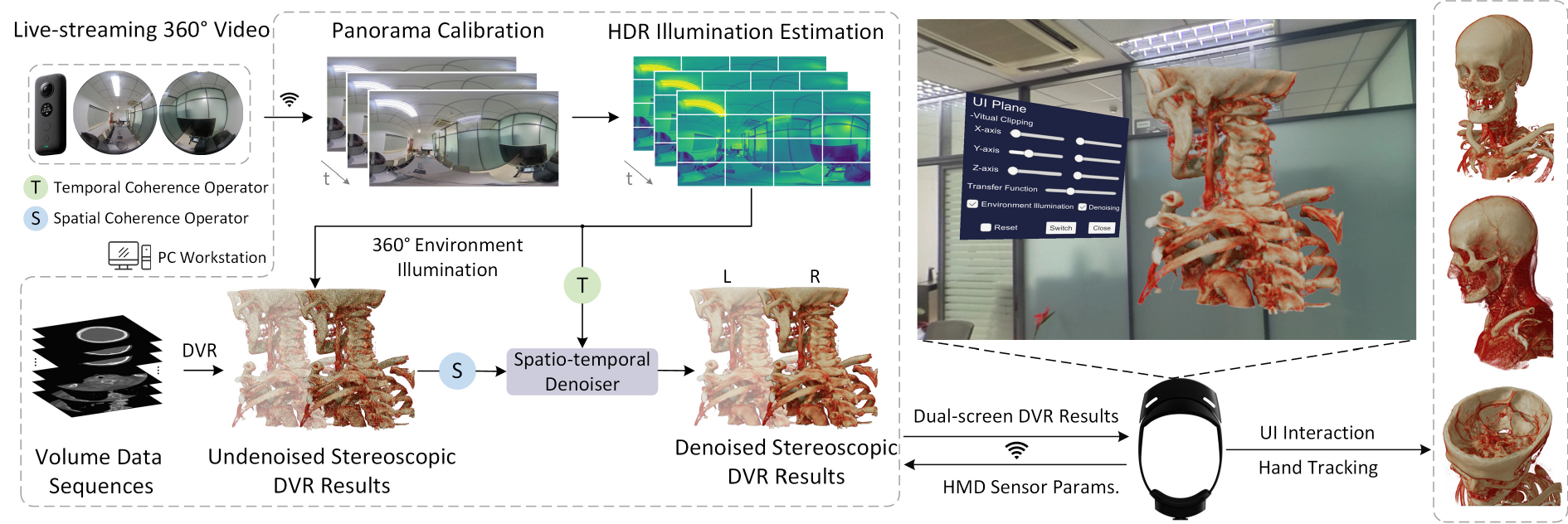}
\vspace{-0.20in}
\caption{Workflow of our MR visualization framework for realistic DVR with environment-synced illumination. Photos of environment scenes are first captured using a dual-fisheye camera and are transmitted to a PC via WiFi. The PC serves as a graphics workstation to perform panorama generation, HDR illumination estimation and realistic DVR with spatio-temporal denoising. Then, DVR results are consecutively transmitted from the PC to an MR HMD via WiFi. The user can inspect and manipulate the display of volumetric data using an MR HMD.}
\label{fig:Totalpipeline}
\vspace{-0.10in}
\end{figure*}

In this paper, we propose a practical MR visualization framework that can produce realistic DVR results with environment-synced illumination in real-time (Fig. \ref{fig:Totalpipeline}). The dynamic HDR environment illumination is first estimated consecutively from panoramic images captured using an LDR panorama camera in real-time. Then, an MC-based VPT algorithm with limited light samples is used to perform realistic DVR. The rendering quality can be greatly optimized by our designed spatio-temporal denoising algorithm, especially for MR development using an HMD. The main contributions of our research work are outlined as follows:

\begin{itemize}

\item We develop a new MR visualization framework that can perform high-quality DVR. The rendered $3$D volumetric structures can synchronously reflect the dynamic environment illumination of real-world scenes in real-time to improve user's sense of immersion during MR viewing.

\item We propose a new spatio-temporal denoising method especially designed for MC-based VPT on an MR HMD. The denoiser is developed to make full use of the DVR results between adjacent screens of an HMD and adjacent frames to optimize the DVR results with panoramic HDR illumination. 

\item We conducted a user study to evaluate advanced volume illumination models in MR. The experiment results demonstrated that the volume visualization in MR can greatly benefit from the environment illumination and shadowing effect.

\end{itemize}

\section{Related Work}

In this section, previous research works on improving the quality of realistic rendering and MR displaying are outlined.

\subsection{MR Rendering Using HMD Devices}

With the rapid development of computer graphics in entertainment, a majority of existing MR applications focus on rendering 3D surface models. Rhee et al. \cite{Rhee2020} presented a telecollaboration platform to teleport virtual surface models into an MR space with visual and audio cues. Eom et al. \cite{NeuroLens2022} designed an optical marker-based system for intraoperative use in neurosurgery. To accelerate graphics computation, foveated rendering system \cite{Guenter2012} was developed to reduce the image resolution of non-foveated regions. Fink et al. \cite{fink2019hybrid} effectively used the relationship between observation distance and binocular disparities to reduce the number of rendered primitives. 

To strengthen user's sense of immersion in displaying volumetric data using an HMD, Scholl et al. \cite{Scholl2018direct} suggested that the update rate of DVR should exceed 90Hz. They reduced the sampling rate and volumetric data size to satisfy this requirement. Besides, Jung et al. \cite{Jung2022} discussed the effect of different sampling distances and mipmapping settings on the computational efficiency of DVR. Waschk et al. \cite{Waschk2020favr} utilized HMD lens distortion and the mechanism of human visual perception to perform adaptive DVR, which reduces the rendering cost. These DVR methods were based on the non-realistic rendering strategy and cannot guarantee the high-fidelity of DVR in MR.

\subsection{Advanced Volume Illumination Techniques}

Advanced volume illumination techniques have been proven to affect user's spatial perception \cite{lindemann2011influence}. Previous researchers found that shadowing effect \cite{Tobias07} and ambient occlusion \cite{ropinski2008interactive} help improve the visual effects of volumetric data. 
Realistic DVR quality can be further improved by applying the physically-based rendering models \cite{Kroes2012interactive, Dappa2016cinematic, engel2016real}. Mixing various bidirectional scattering distribution functions (BSDF) can facilitate approximating different physically real materials \cite{Pharr16}. Kroes et al. \cite{Kroes2012interactive} presented a GPU-based interactive DVR renderer that integrates stochastic ray-traced lighting to improve the realisticness of DVR. 


Benefiting from the development of panorama photography techniques, an omnidirectional radiance map can provide natural HDR environment illumination \cite{Pharr16}. Some research works have applied the HDR radiance map to DVR. Volumetric spherical harmonic illumination \cite{khlebnikov2014parallel} encoded light visibility and HDR radiance maps to achieve low-frequency illumination in real-time DVR. Zhang et al. \cite{Zhang2013} combined volumetric photon mapping and virtual point lights (VPL) technique for global indirect illumination. Von et al. \cite{Von2016Efficient} presented an approximate light visibility structure and performed joint importance sampling, which accelerated the convergence rate of DVR results. However, the DVR speeds of these realistic solutions are hard to satisfy real-time MR applications.

\section{Interactive Realistic DVR with Dynamic Environment Illumination}
\label{sec:DVRandlighting}

To deploy a practical realistic DVR on an MR HMD, both accurate and efficient environment illumination estimation and realistic DVR are indispensable for enabling real-time MR displaying.

\subsection{Reconstruction of Environment-Synced Illumination}
\label{subsec:illumination_estimation}

The initial task is to real-time obtain stitched images from a panorama camera. We directly encode raw fisheye images into low-latency H.264 format and transmit it to a PC workstation. Then, an efficient panorama reconstruction algorithm \cite{Cheng2022CGF} is used to create high-quality panoramic images. In our MR system, a dual-lens fisheye camera is attached to an MR HMD, so it always moves together with the MR HMD. They both can be allocated in the same coordinate system.

To align the environment light with the rendering target in the same coordinate system, we designed a spatial calibration operator for panorama reconstruction. The initial stitched panoramic image $I_{c}$ in the camera coordinate system is first converted to its version $I_{w}$ in the world coordinate system. The transformation matrix of the current user pose relative to the initial user pose can be calculated based on HMD sensor parameters. When a user moves volumetric data, the relative position between the user and volumetric data would change accordingly. The appearance of rendered volumetric structures should exhibit spatially-varying illumination effects, especially in indoor scenes \cite{Gardner17}. $I_{w}$ does not include depth information and is typically mapped onto a $3$D unit sphere as the radiance map. To provide more accurate illumination of the rendered target, $I_{w}$ can be further reconstructed to a warped panoramic image $I^{'}_{w}$ centered at volumetric data \cite{Gardner17}. 

Since $I^{'}_{w}$ is an LDR image with a limited pixel bit depth, important lighting information is usually discarded. To obtain real-world illumination from $I^{'}_{w}$, we first perform the light source detection and lighting intensity calculation from the real-time illumination estimation method \cite{Cheng2022} to predict the HDR radiance map $Q$. The variation of light conditions in adjacent frames can be used to guide the temporal denoising weights of rendering results. A simple way to quantify the illumination difference is to calculate the average illumination difference. However, when a bright light source quickly moves in a scene, although the shading effect of rendered models usually significantly changes, the average illumination difference between adjacent frames is subtle. For this reason, $Q$ is divided into multiple $N\times N$ image patches. The illumination differences of the corresponding patches between adjacent frames are calculated separately. When the illumination of any patch varies greatly, it is considered that the illumination of the whole scene varies greatly. The illumination similarity of $k$-th patch between $Q_{t-1}^{k}$ and $Q_{t}^{k}$ can be calculated using log-SSIM \cite{valenzise2014} and is defined as $\tau (Q_{t-1}^{k},Q_{t}^{k})$. The illumination difference between adjacent frames can be calculated as:


\vspace{-0.10in}
\begin{equation}\label{eq:lightdifference}
  T = 1 - \max_{k\in N^{2}} \left(\tau (Q_{t-1}^{k},Q_{t}^{k})\right),
\end{equation}
$T\in[0,1)$ is used for subsequent DVR. The closer $T$ is to $0$, the smaller the illumination difference between adjacent frames. The proposed DVR method employs the VPT algorithm based on the MC strategy according to its vertical ordinate \cite{Pharr16}.


\subsection{Effective Realistic DVR}
\label{subsec:effective_DVR}

To improve realisticness and support more advanced illumination, an MC-based VPT algorithm \cite{Kroes2012interactive,Von2016Efficient} is adopted in our method to estimate the light transportation with a single scattering effect. In real-time rendering, limited shading samples inevitably cause lots of noise. To solve the conflict between the result quality and computational efficiency of VPT-based DVR, we develop a real-time denoising algorithm, especially for interactive MR viewing.

\vspace{-0.05in}
\subsubsection{Reprojection Strategy for MR HMD}
\label{subsubsec:Reprojection}

The stereoscopic effect perceived by a user is the combination of the two screen images reflected in the human brain. Adjacent HMD screens and adjacent frames have high spatial and temporal coherence respectively. Generally, the rendering noise distribution of the two screen images is inconsistent. This aggravates the visual discomfort perceived by human eyes. 

\begin{figure}[h]
  \centering 
  \begin{overpic}[width=0.7\columnwidth]{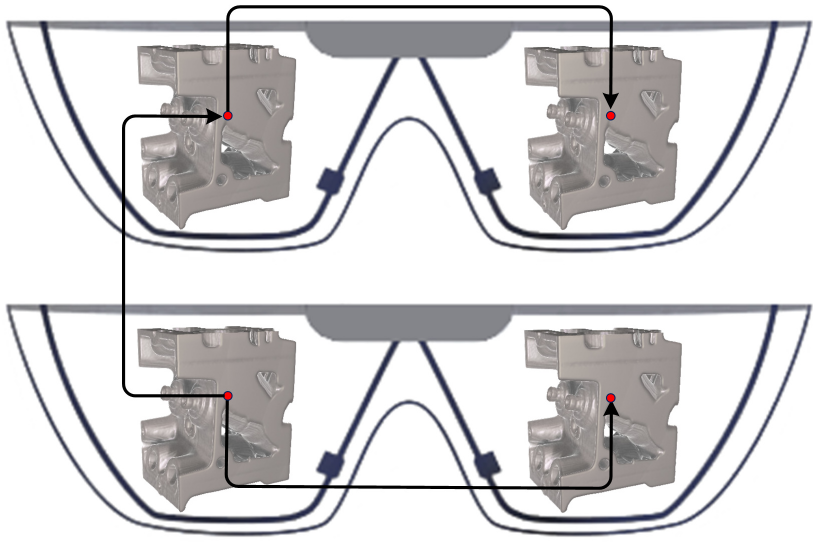}
    \put(-5,38){\scriptsize\rotatebox{90}{Frame $t-1$}}
    \put(-5,8){\scriptsize\rotatebox{90}{Frame $t$}}
    \put(16.5,31.2){\scriptsize$\pi_{L}^{t-1}(\cdot)$}
    \put(44,68){\scriptsize$\pi_{R}^{t-1}(\cdot)$}
    \put(45,2.2){\scriptsize$\pi_{R}^{t}(\cdot)$}
    \put(28,16){\scriptsize$k_{L}^{t}$}
    \put(28,48.5){\scriptsize$k_{L}^{t-1}$}
    \put(76,16){\scriptsize$k_{R}^{t}$}
    \put(76,48.5){\scriptsize$k_{R}^{t-1}$}
  \end{overpic} 
  \caption{Taking the left screen of an HMD as an example, the reprojection strategy is implemented between the two HMD screens and adjacent DVR result frames.}
  \label{fig:sample}
  \vspace{-0.10in}
\end{figure}

To improve the quality of DVR results, we used the reprojection strategy \cite{Diego2007} and optimized it for realistic DVR in an MR HMD. The new reprojection strategy aims to exploit temporal coherence between adjacent frames and spatial coherence between the two HMD screens. Specifically, the left and right screens of an HMD are marked as $L$ and $R$ respectively. The $k$-th pixel on the left screen in the $t$-th frame is marked as $k_{L}^{t}$. Its first scattering position nearest to the virtual camera in $3$D virtual space \cite{Iglesias2020real} is labeled as $\mathbf{x}(k_{L}^{t})$. $\pi(\cdot)$ indicates the $2$D projection operator to reproject pixel from $3$D virtual space. Taking the pixel $k_{L}^{t}$ as an example, the reprojection strategy can be implemented on the two HMD screens and adjacent frames. Four samples can thus be obtained at once, as shown in Fig.\ref{fig:sample}. The corresponding reprojected pixels of pixel $k_{L}^{t}$ can be represented as:

\vspace{-0.10in}
\begin{equation}
   k_{L}^{t-1} \leftarrow \pi_{L}^{t-1}(\mathbf{x}(k_{L}^{t})),  k_{R}^{t-1} \leftarrow \pi_{R}^{t-1}(\mathbf{x}(k_{L}^{t-1})),  k_{R}^{t} \leftarrow \pi_{R}^{t}(\mathbf{x}(k_{L}^{t})).   
\end{equation}

\begin{figure}[h]
  \centering 
  \begin{overpic}[width=0.95\columnwidth]{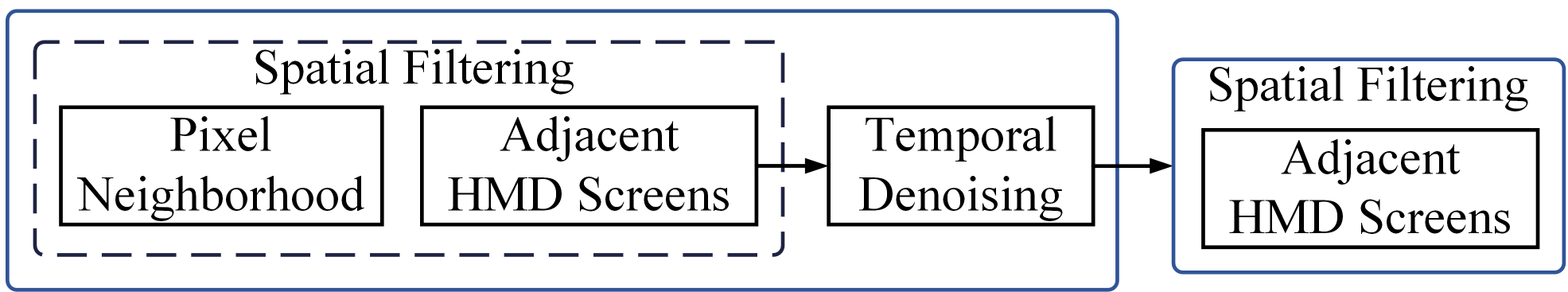}
    \put(32,19.5){$\mathcal{D}_{1}$}
    \put(85,16.5){$\mathcal{D}_{2}$}
  \end{overpic} 
  \caption{Overview of our spatio-temporal denoising method. It consists of two steps and reuses the information of pixel neighborhoods, adjacent HMD screens and adjacent frames.}
  \label{fig:denoisingPipline}
  \vspace{-0.10in}
\end{figure}

\vspace{-0.05in}
\subsubsection{Spatio-temporal Denoising}
\label{subsubsec:denoiser}

The proposed spatio-temporal denoising method is designed as a two-step process (Fig. \ref{fig:denoisingPipline}). The first step is labeled as $\mathcal{D}_{1}$. Images in two HMD screens are denoised using the pixel information of an image neighborhood area, previous frames and sample coherence between two HMD screens. To improve DVR speed, each pixel in an image is sampled by two rays in our renderer. Therefore, eight samples in each frame participate using our optimized reprojection strategy. Since the stable surface intersection between the ray and volumetric data in VPT generally does not exist \cite{Iglesias2020real}, it is hard to guarantee a one-to-one correspondence between the reprojected pixels and therefore can affect the quality of denoised results. After $\mathcal{D}_{1}$, the DVR results may still have slight noise and disturb the displaying in the two HMD screens. Then, the second step labeled as $\mathcal{D}_{2}$ is applied to further eliminate these inter-screen noise and improve the user's visual comfort during MR viewing.

$v(k_{L}^{t})$ represents the initial pixel value of $k_{L}^{t}$. $v(k_{L}^{t},1)$ and $v(k_{L}^{t},2)$ denote the denoised results obtained after $\mathcal{D}_{1}$ and $\mathcal{D}_{2}$ respectively. Taking $v(k_{L}^{t})$ as an example, the denoised pixel values obtained after each iteration step can be represented as: 

\vspace{-0.10in}
\begin{equation}
  \begin{split}
v(k_{L}^{t},1) & = \mathcal{D}_{1} 
  \begin{pmatrix} v(k_{L}^{t}), v(k_{R}^{t}), v(k_{L}^{t-1}, 2) \end{pmatrix}
  \\
v(k_{L}^{t}, 2) & = \mathcal{D}_{2} \big(  v(k_{L}^{t},1), v(k_{R}^{t},1)  \big)
  \end{split},
\end{equation}

{\noindent}where $v(k_{L}^{t},2)$ is the final result of denoising pixel $k_{L}^{t}$. $v(k_{L}^{t-1},2)$ indicates the previous sample in $(t-1)$-th frame. As shown in Fig. \ref{fig:sample} and Fig. \ref{fig:denoisingPipline}, $\mathcal{D}_{1}$ first uses a spatial filtering on the neighborhood of a pixel per image and edge-preserving bilateral filtering on corresponding pixel samples of $v(k_{L}^{t})$ and $v(k_{R}^{t})$ on the two HMD screens. The bilateral weight is calculated according to the albedo, the gradient at $\mathbf{x}(\cdot)$ and $z$-depth of $\mathbf{x}(\cdot)$ between a pixel and its reprojected pixels. The output result after bilateral filtering is expressed as $\widetilde{v}(k_{L}^{t}, 1)$. Then, the previous samples are used for the temporal denoising $\widetilde{v}(k_{L}^{t}, 1)$. The illumination difference $T$ (as described in Eq. (\ref{eq:lightdifference})) is used to calculate the weight of the previous sample. The weight of $\widetilde{v}(k_{L}^{t}, 1)$ is equal to $1$ and the weight of denoised pixel in the previous frame $v(k_{L}^{t-1},2)$ is calculated as:

\vspace{-0.10in}
\begin{equation}
  w_{v} = \delta(\frac{1}{T - 1})\cdot \zeta,
\end{equation}

{\noindent}where $\delta(\cdot)$ is a sigmoid function. A higher $T$ corresponds to a lower output value of $\delta$. It indicates that the rendering result in the previous frame has a small contribution to the current rendering result. $\zeta$ indicates the bilateral weight calculation between adjacent frames. After the process of $\mathcal{D}_{1}$ filtering, $\mathcal{D}_{2}$ filtering is then performed between the two adjacent HMD screens:

\vspace{-0.15in}
\begin{equation}
  \begin{split}
  v(k_{L}^{t},2) = \lambda  \cdot v(k_{L}^{t},1) + (1 - \lambda)\cdot v(k_{R}^{t},1)\\
  \lambda  = \alpha \cdot (\beta  - \exp^{-1}(||\gamma(k_{L}^{t}) - \gamma(k_{R}^{t})  ||))
  \end{split},
  \vspace{-0.10in}
\end{equation}

{\noindent}where $\lambda$ is the weight factor that controls the contribution proportion between pixel values in the left and right screens. $\gamma(\cdot)$ denotes the albedo of corresponding pixels. In our implementation, the factors $\alpha$ and $\beta$ are set to $0.5$ and $2$ respectively. In this case, if the difference between $\gamma(k_{L}^{t})$ and $\gamma(k_{R}^{t})$ is small, $\lambda$ is close to $0.5$, so that two screens have the same contribution. In contrast, if $\gamma(k_{L}^{t})$ and $\gamma(k_{R}^{t})$ are quite different, $v(k_{L}^{t},2)$ relies more on $v(k_{L}^{t},1)$. After the two-step denoising process, the final dual-screen DVR images are transmitted to an MR HMD via WiFi.

\begin{figure*}[t]
  \centering
\hspace{-0.10in}
\subfloat[Li et al. \cite{Li20213d}\label{subfig:f0}]{
\begin{minipage}[b]{0.24\textwidth}
\centering
\includegraphics[width=\textwidth]{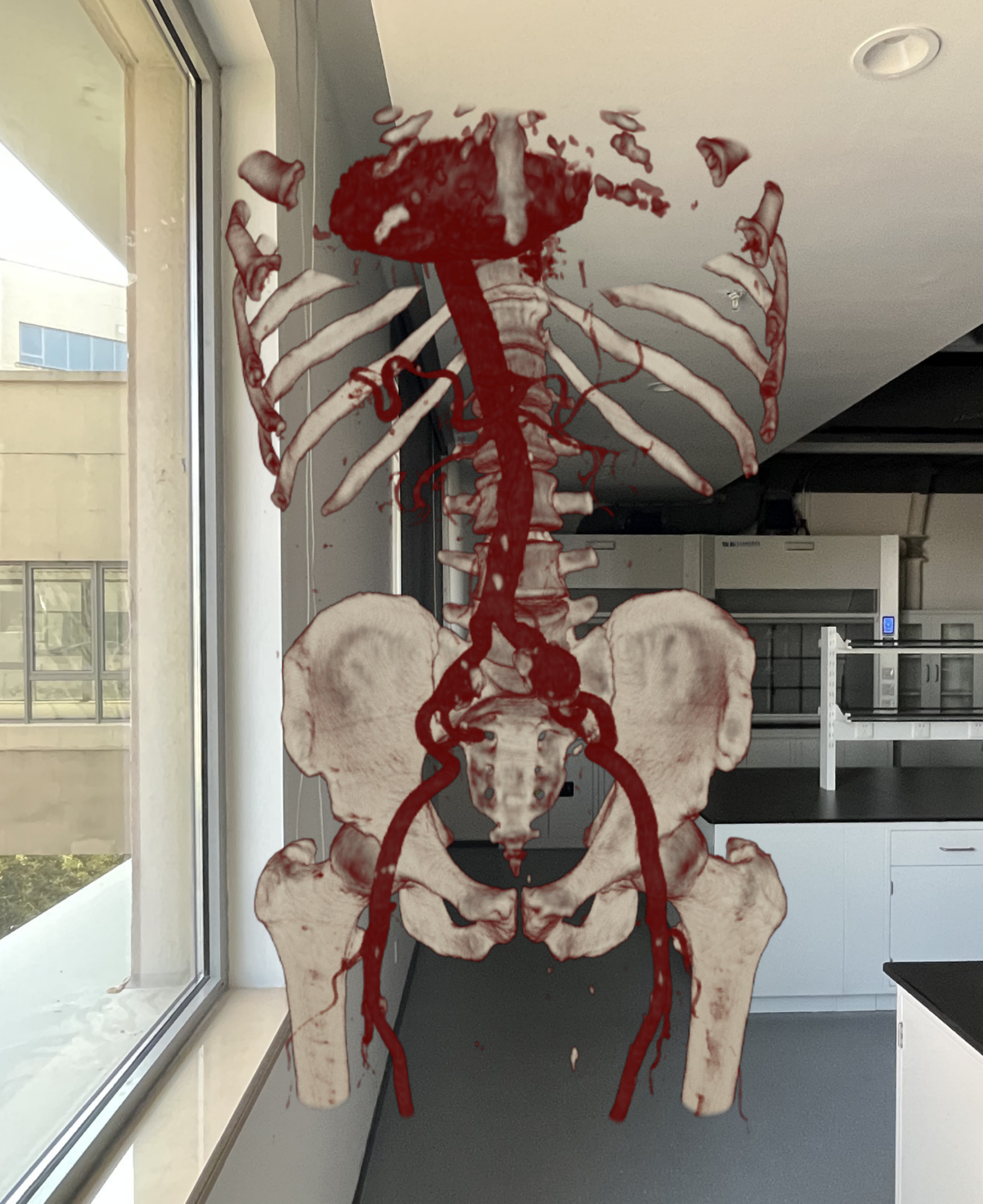}
\end{minipage}
}  \hspace{-0.05in}
\subfloat[Waschk et al. \cite{Waschk2020favr}\label{subfig:f3}]{
\begin{minipage}[b]{0.24\textwidth}
\centering
\includegraphics[width=\textwidth]{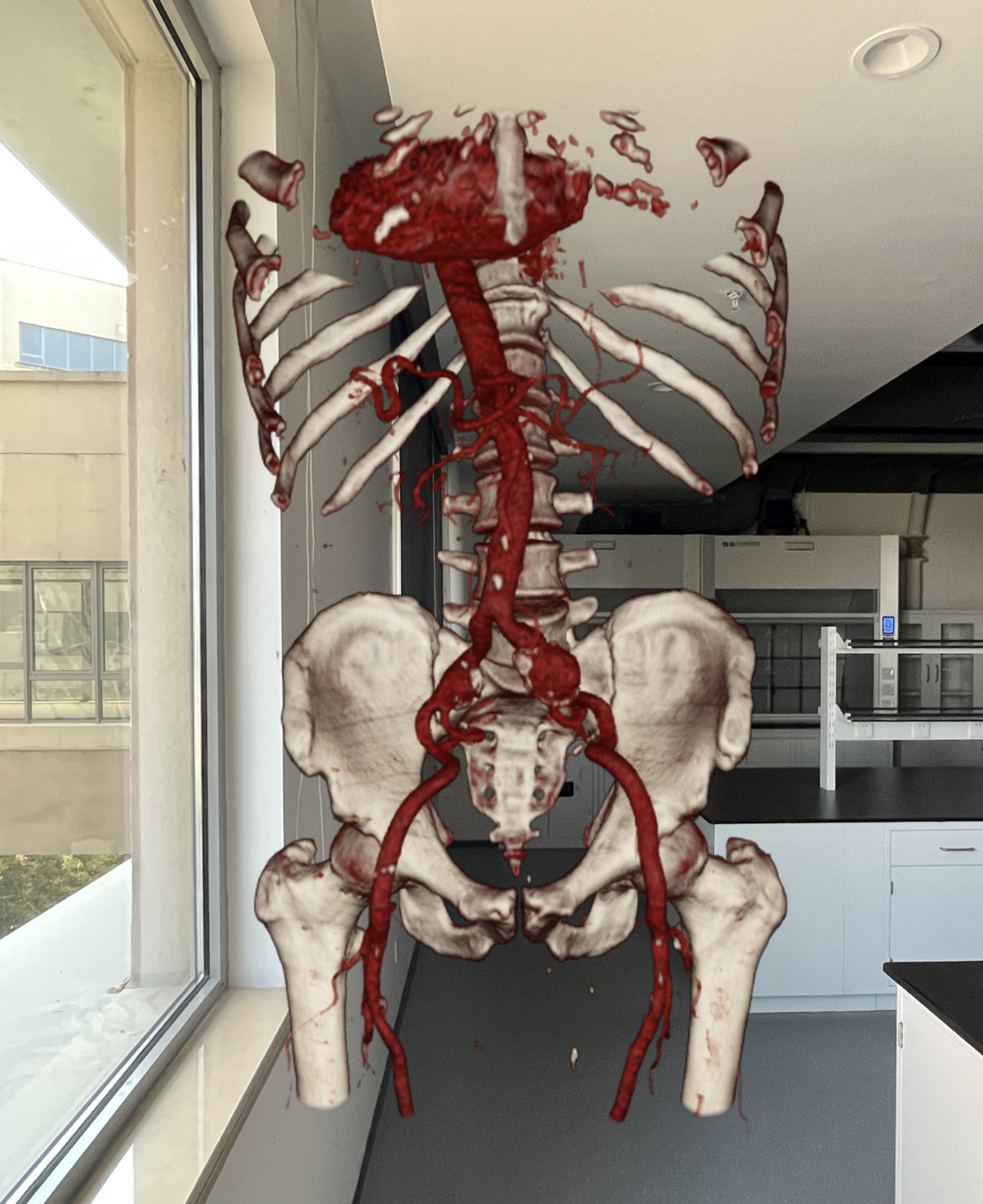}
\end{minipage}
} \hspace{-0.05in}
\subfloat[Rhee et al. \cite{Rhee2020}\label{subfig:f4}]{
\begin{minipage}[b]{0.24\textwidth}
\centering
\includegraphics[width=\textwidth]{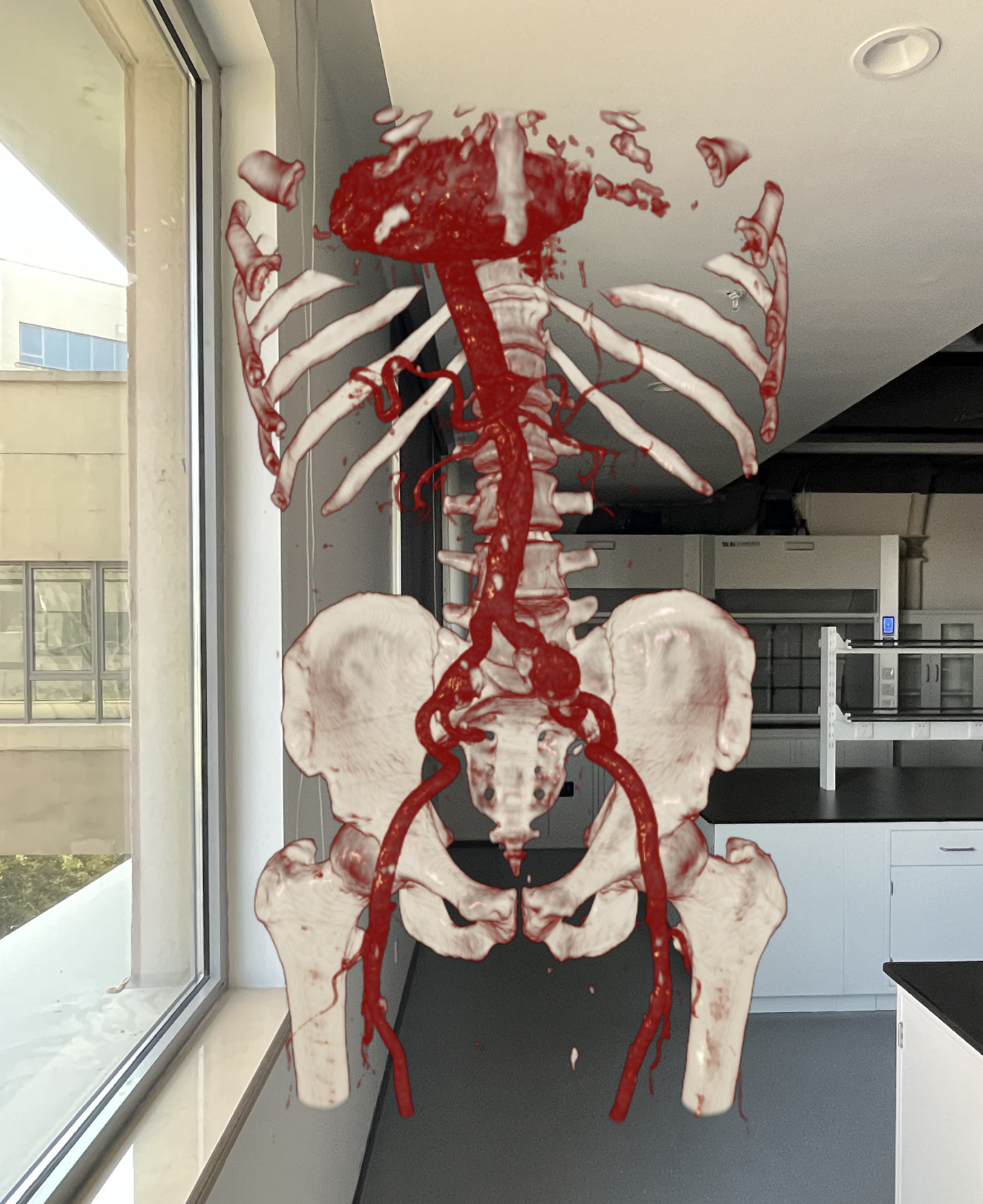}
\end{minipage}
} \hspace{-0.05in}
\subfloat[The proposed method\label{subfig:f4}]{
\begin{minipage}[b]{0.24\textwidth}
\centering
\includegraphics[width=\textwidth]{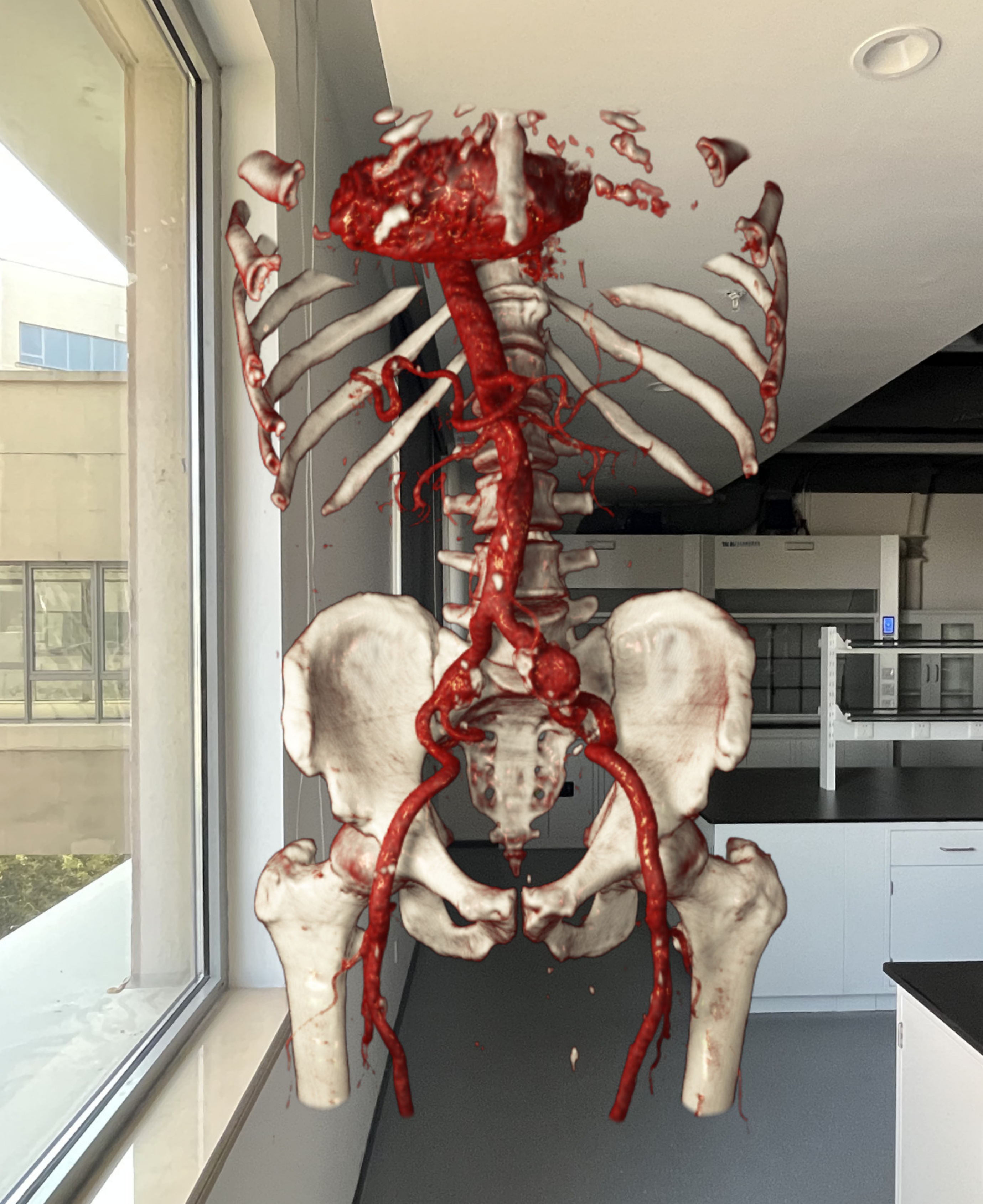}
\end{minipage}
}
\vspace{-0.10in}
\caption{Visual comparison of rendering results generated using different volume illumination techniques in MR to verify the impact of environment illumination and shadowing effect.}
\label{fig:keyframes}
\vspace{-0.10in}
\end{figure*}

\section{Evaluations}
\label{sec:Evaluations}

The proposed MR visualization framework was mainly executed on a PC workstation that has an Intel i7-12700KF CPU with $32$GB RAM and an NVIDIA GeForce RTX $3080$ Ti GPU with $12$GB video memory. An Insta360 ONE X camera was used to capture fisheye images of surrounding environment scenes. The resolution of the reconstructed panoramic image was set to $512 \times 256$ to accelerate illumination estimation. Microsoft HoloLens 2 was selected as the HMD device for MR displaying ($1440 \times 936$ per eye).

\vspace{-0.05in}
\subsection{User Study Design}
\label{subsec:User}

The user study aimed to investigate the effect of different volume illumination strategies on the user's MR experiences.

\vspace{-0.05in}
\subsubsection{Evaluated Volume Illumination Techniques}
\label{subsubsec:Studymethods}

Three additional volume illumination techniques that have been used in VR, AR and MR applications were reproduced for comparison with our method.

\textbf{Absorption-emission model} was utilized by Li et al. \cite{Li20213d} to visualize $3$D medical data in their AR system. Absorption-emission model is the most common and lightweight volume visualization model \cite{salama2007gpu}. It simulates the self-luminescence of volume particles without considering the light scattering. 

\textbf{Gradient-based shading model} was optimized by Waschk et al. \cite{Waschk2020favr} to implement the rendering effect in their VR system. This model can enhance surface details by the Phong model \cite{levoy1988display}. This model was included to mainly compare the visual quality with and without global shadows.



\textbf{Pre-filtered environment mapping model} was used by Rhee et al. \cite{Rhee2020} in their MR system. They used a pre-filtered cubemap with multiple mipmap levels for image-based lighting (IBL). We added this method to compare the shading effect using their LDR illumination and our HDR illumination.

\subsubsection{Participant Selection and Questionnaire}
\label{subsubsec:StudySurvey}

A participant with prior knowledge of anatomy can have difficulties to objectively assess the depth and shape information of rendered 3D structures in volume visualization. We try to investigate the optimal visualization based on the assessment given by the users without professional medical knowledge (e.g. general patients and non-medical students). In our user study, $25$ participants ($9$ females and $16$ males) aged from $20$ to $50$ years old with an average of $26.8$ years were invited. Before this study, a pre-experiment survey was given to each participant to ask demographic questions about prior experience in MR. Results showed that $4$ participants had experience in MR development and always used MR HMDs, $9$ participants had worn an MR HMD once or twice, and $12$ participants had never worn an MR HMD before. None of the participants had experience with DVR development.

\begin{table}[h]
\centering
\vspace{-0.10in}
\caption{Post-experiment Questionnaire.}
\vspace{-0.10in}
\renewcommand{\arraystretch}{1}
\setlength\tabcolsep{1pt}
\begin{tabular}{C{8mm} | p{1mm} p{72mm}}
\hlineB{2}
  &  &\multicolumn{1}{c}{Questions}\\
\hline
Q1 & &The rendering results of volumetric data (e.g., material appearances and shading effects) were realistic in MR. \\
Q2 & &The rendering results were free of visible noise and ghost artifacts.\\
\hline
Q3 & &This volume illumination technique was beneficial for observing the depth and shape of volumetric structures.\\
Q4 & &The volumetric data can blend naturally with real-world scenes.\\
\hline
Q5 & &The MR displaying was robust and showed no visible drift or jittering effects. \\
Q6 & &The user can efficiently manipulate the volumetric data via hands or UI modules. \\
\hline
Q7 & &Please write any other comments or feedback about your MR experience.\\
\hlineB{2}
\end{tabular}
\vspace{-0.05in}
\label{tab:questions}
\end{table}

All volumetric data were randomly ordered for each participant. Participants can freely switch between different volume illumination models and volumetric data. To control the individual bias of participants, we designed several questions in different categories to ensure that participants made objective and fair assessments (Table \ref{tab:questions}). Three categories in post-experiment survey are: \emph{realistic DVR quality} (Q1 - Q2), \emph{user perception of MR content} (Q3 - Q4) and \emph{MR interaction experience} (Q5 - Q6). At the end of the questionnaire, we asked participants to leave any open-ended feedback about MR experience (Q7). After inspecting and manipulating MR rendering results for ten minutes, each participant was then asked to respond to each question in Table \ref{tab:questions} and rate the experience from $1$ to $5$ ($1$ stands for "strongly disagree" and $5$ stands for "strongly agree"). 


\begin{table}[h]
  \footnotesize
\centering
\vspace{-0.05in}
\caption{ANOVA results. ($MS$-within and $MS$-between indicate mean square within and between groups respectively.)}
\vspace{-0.05in}
\renewcommand{\arraystretch}{1.0}
\setlength\tabcolsep{1pt}
\begin{tabular}{C{13mm}| C{16mm} C{16mm} C{16mm} C{16mm}}
\hlineB{2}
Questions & $MS$-within & $MS$-between & $F$-ratio & $p$-value\\
\hline
Q1 & 0.23 & 49.08 & 210.34 & \textbf{\textless 0.001} \\
Q2 & 0.24 & 1.13 & 4.64 & \textbf{0.004} \\
Q3 & 0.22 & 52.65 & 237.53 & \textbf{\textless 0.001} \\
Q4 & 0.37 & 63.21 & 172.78 & \textbf{\textless 0.001} \\
Q5 & 0.61 & 0.09 & 0.148 & 0.93 \\
Q6 & 0.57 & 0.15 & 0.26 & 0.86 \\
\hline
Overall & 1.10 & 64.61 & 58.77 & \textbf{\textless 0.001} \\
\hlineB{2}
\end{tabular}
\label{tab:anova}
\end{table}

\vspace{-0.15in}
\subsubsection{Comparison Results and Discussion}
\label{subsubsec: CompareDifferentDVRMethods}

Statistics were obtained using a one-way analysis of variance (ANOVA) to reject the null hypothesis that all correctness means were equal between techniques. A $p$-value $\leq 0.05$ implies a statistically significant difference between different conditions. The Bonferroni correction for a post-hoc test was then performed to further reveal differences between individual techniques. The ANOVA test in Table \ref{tab:anova} showed a significant difference in the overall MR experience between different techniques ($F(3,596) = 58.77$, $p<0.001$).


\begin{figure}[h]
\centering
\begin{minipage}[b]{0.95\linewidth}
\centering
\hspace{0.08in}
\includegraphics[width=0.312\linewidth]{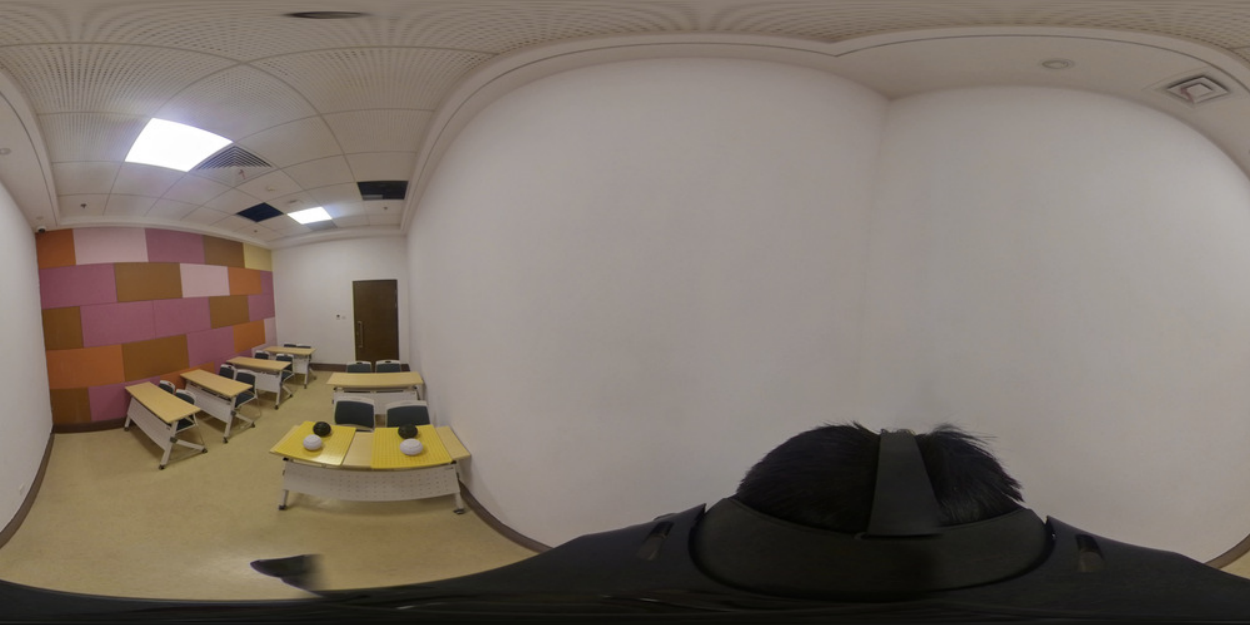}\hspace{-0.01in}
\includegraphics[width=0.312\linewidth]{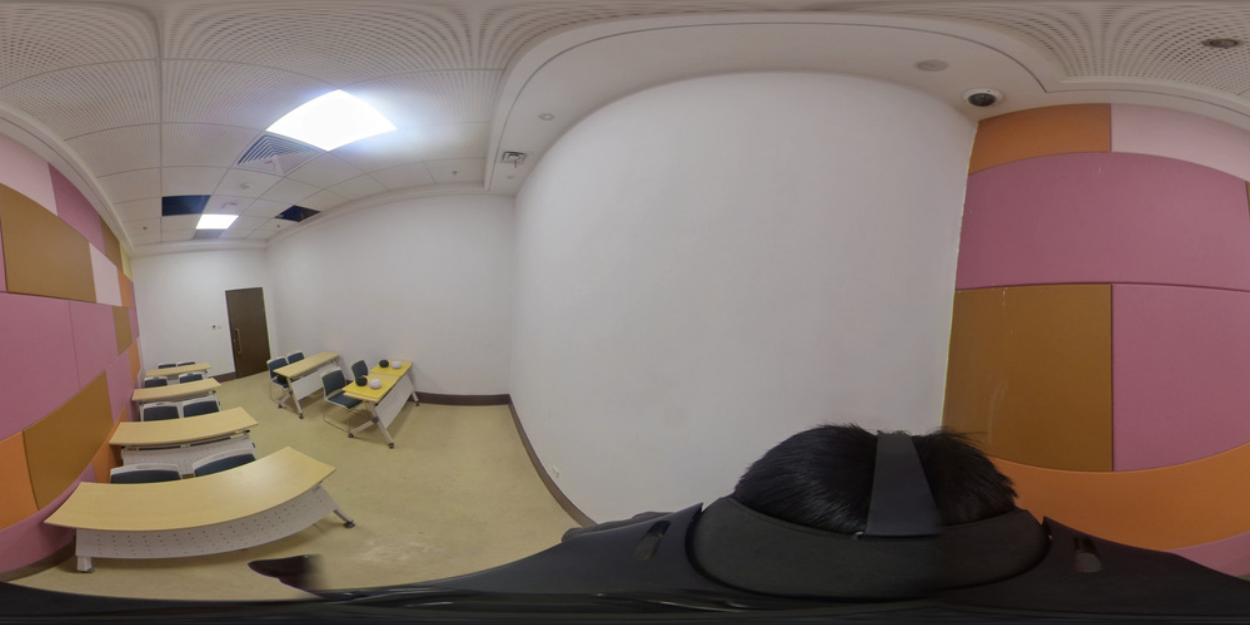}\hspace{-0.01in}
\includegraphics[width=0.312\linewidth]{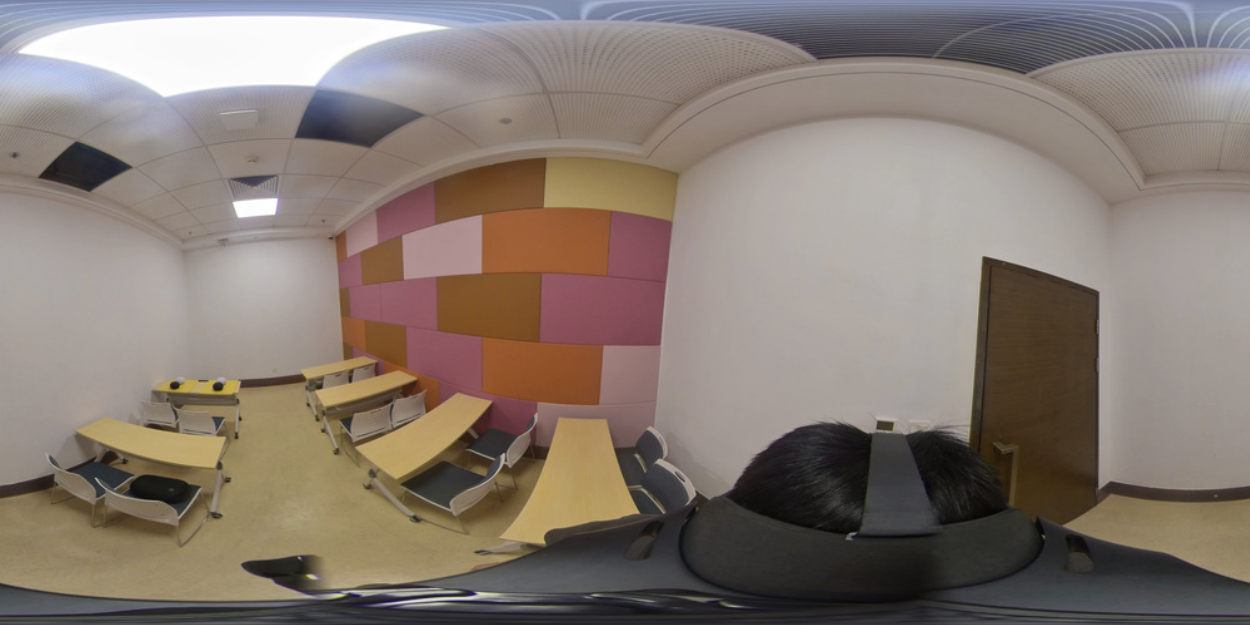}
\end{minipage}\\\vspace{0.02in}
\begin{minipage}[b]{0.95\linewidth}
\centering
\raisebox{0.13in}{\rotatebox{90}{\scalebox{.9}{\scriptsize{Li et al. \cite{Li20213d}}}}}
\includegraphics[width=0.312\linewidth]{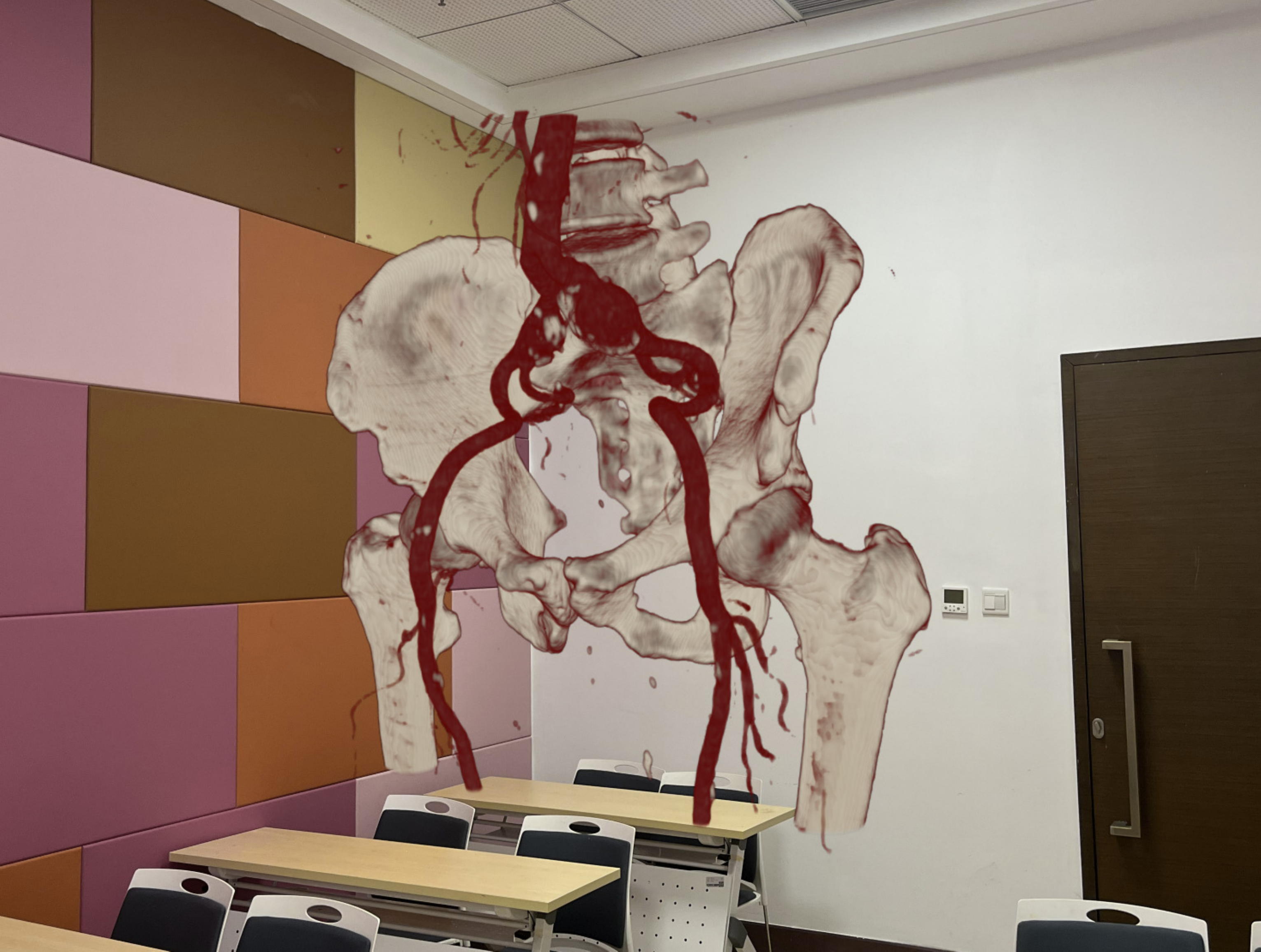}\hspace{-0.01in}
\includegraphics[width=0.312\linewidth]{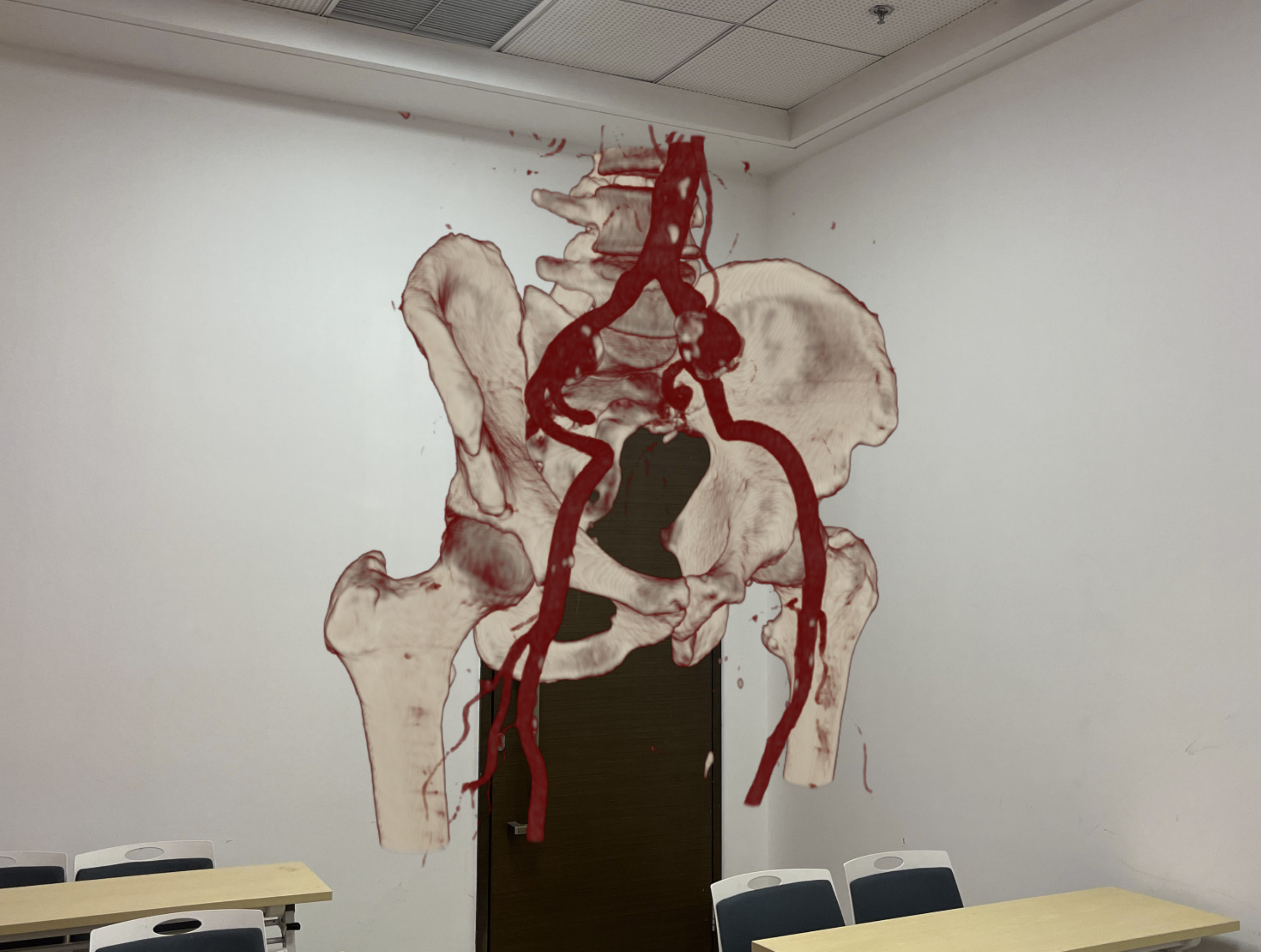}\hspace{-0.01in}
\includegraphics[width=0.312\linewidth]{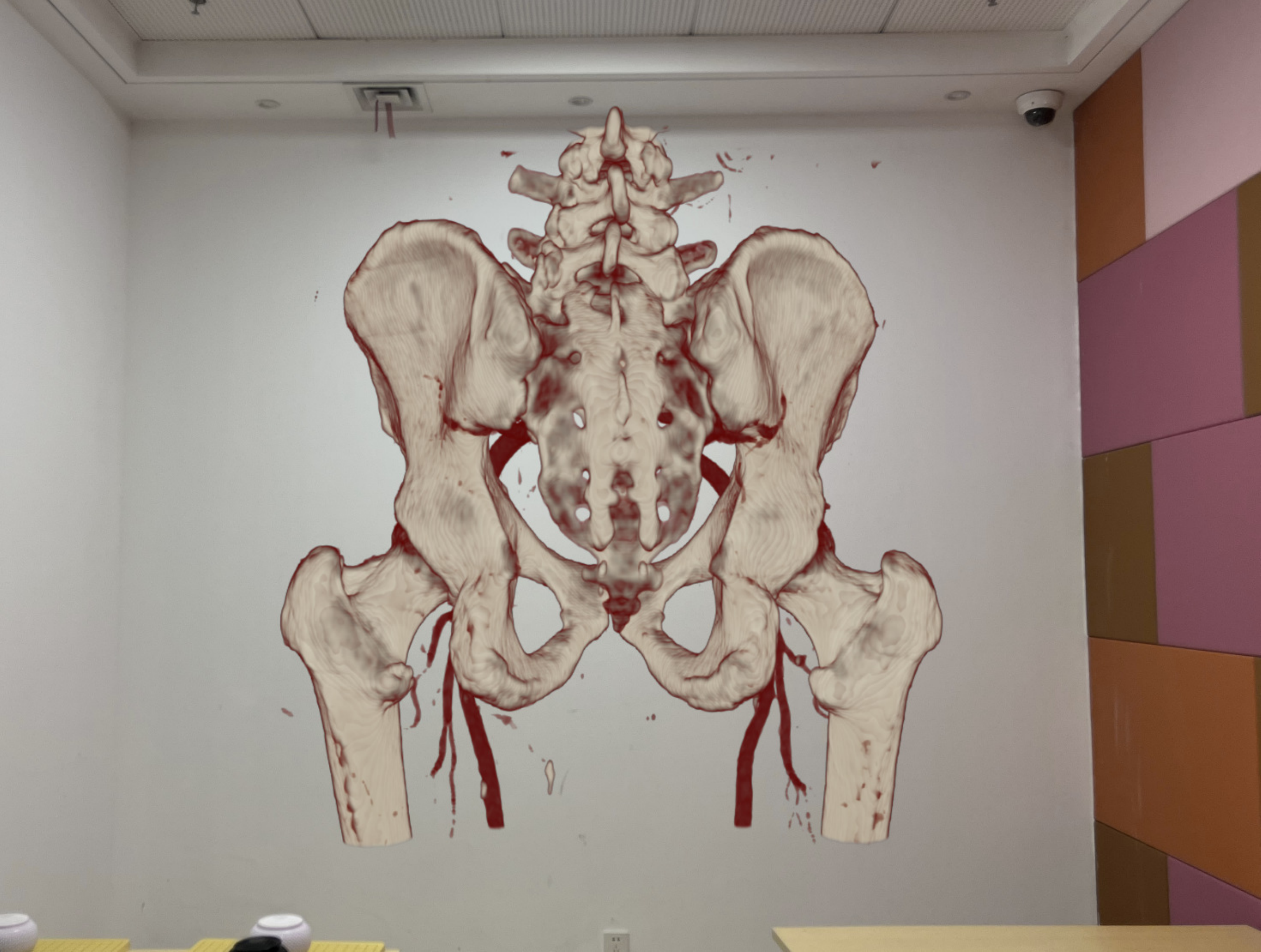}
\end{minipage}\\
\begin{minipage}[b]{0.95\linewidth}
\centering
\raisebox{0.02in}{\rotatebox{90}{\scalebox{.9}{\scriptsize{Waschk et al. \cite{Waschk2020favr}}}}}
\includegraphics[width=0.312\linewidth]{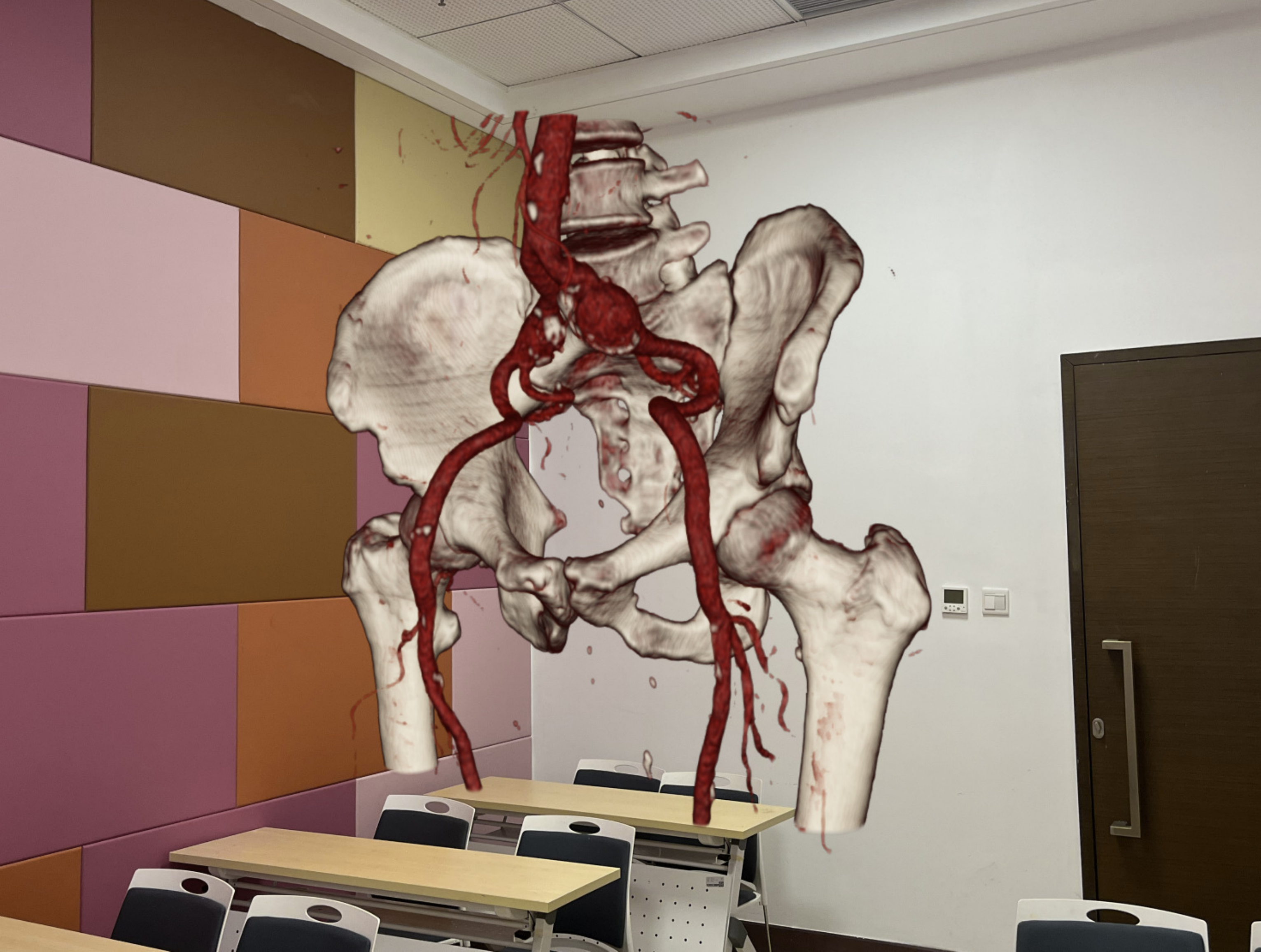}\hspace{-0.01in}
\includegraphics[width=0.312\linewidth]{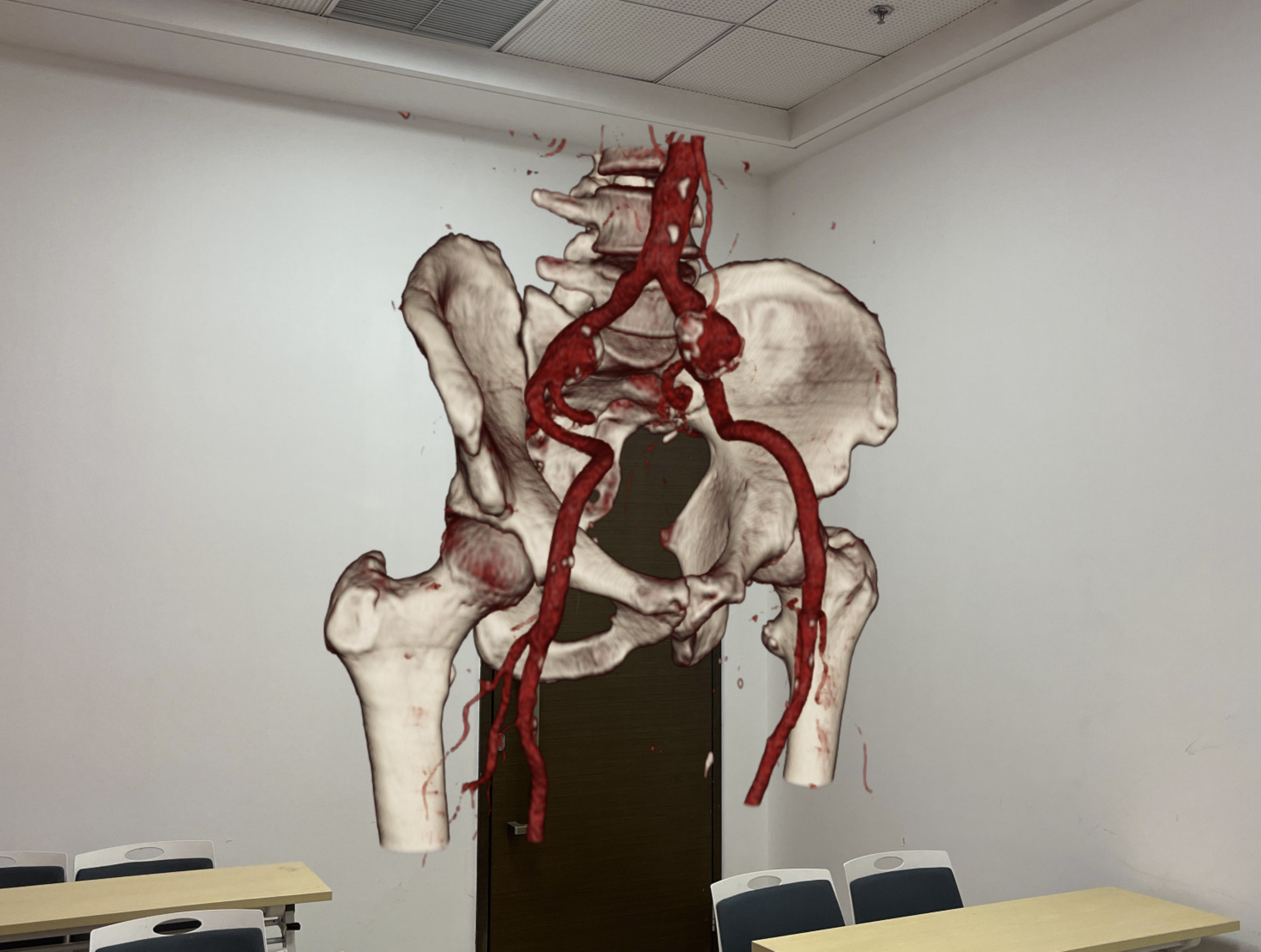}\hspace{-0.01in}
\includegraphics[width=0.312\linewidth]{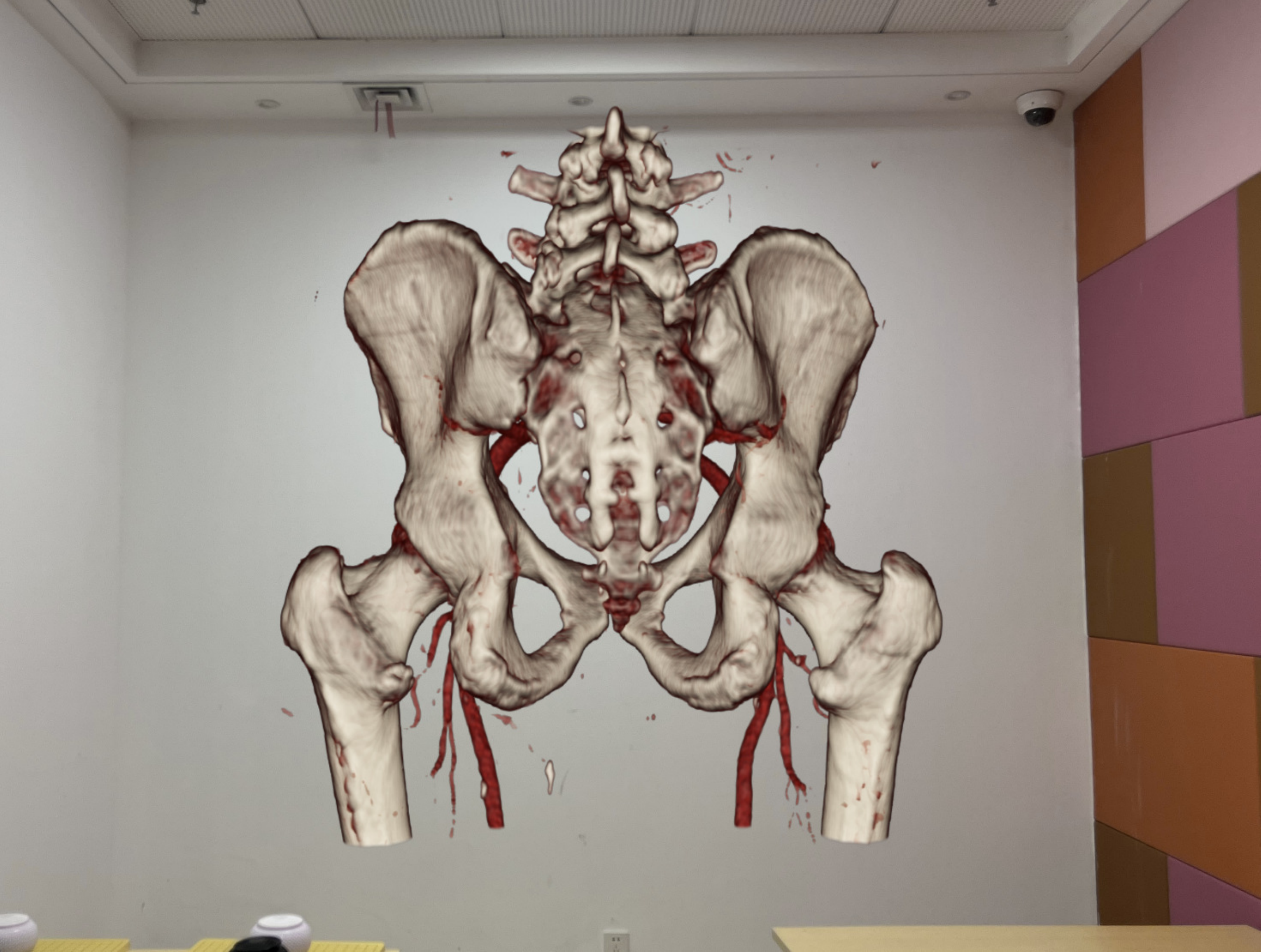}
\end{minipage}\\
\begin{minipage}[b]{0.95\linewidth}
\centering
\raisebox{0.08in}{\rotatebox{90}{\scalebox{.9}{\scriptsize{Rhee et al. \cite{Rhee2020}}}}}
\includegraphics[width=0.312\linewidth]{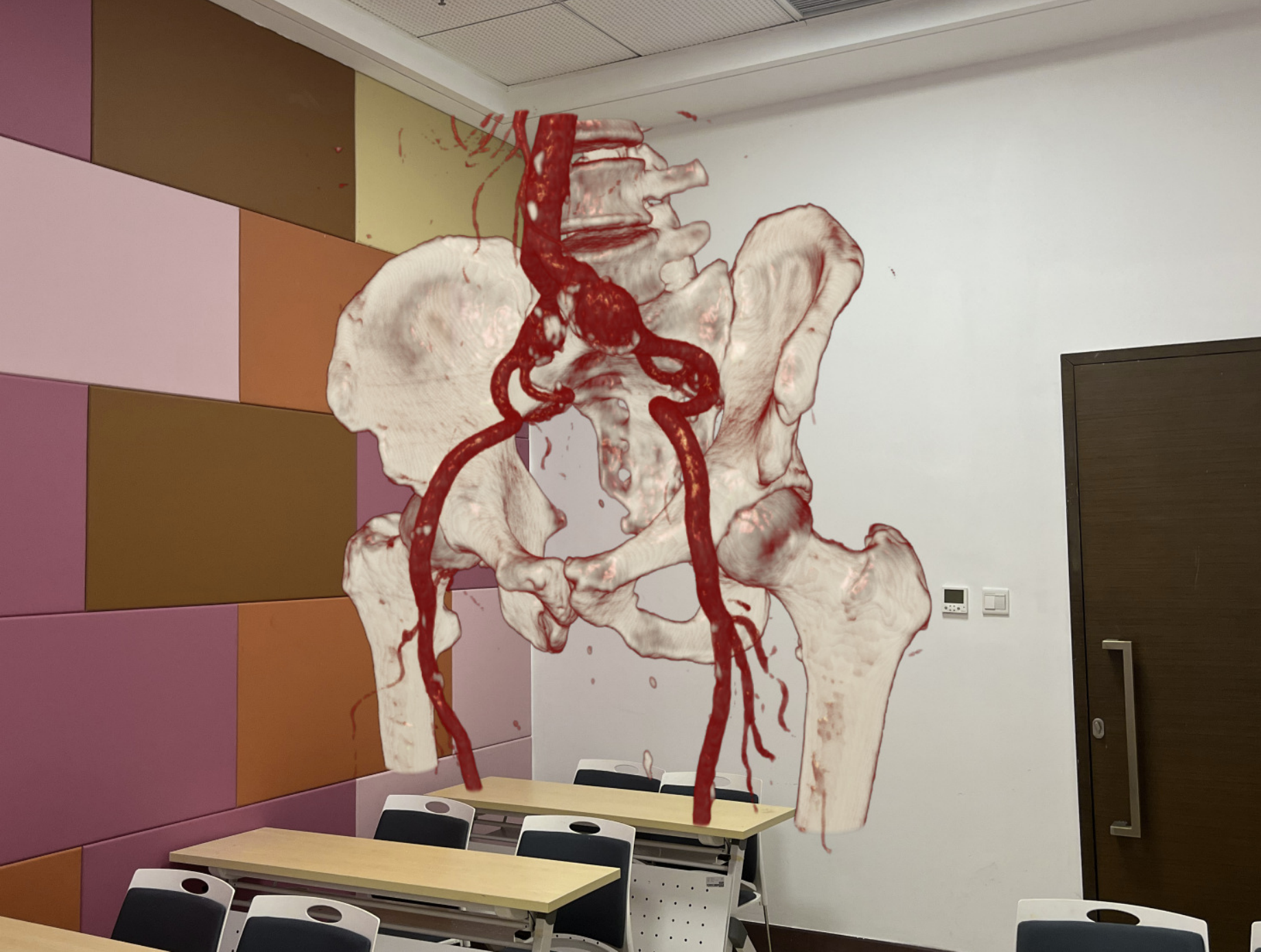}\hspace{-0.01in}
\includegraphics[width=0.312\linewidth]{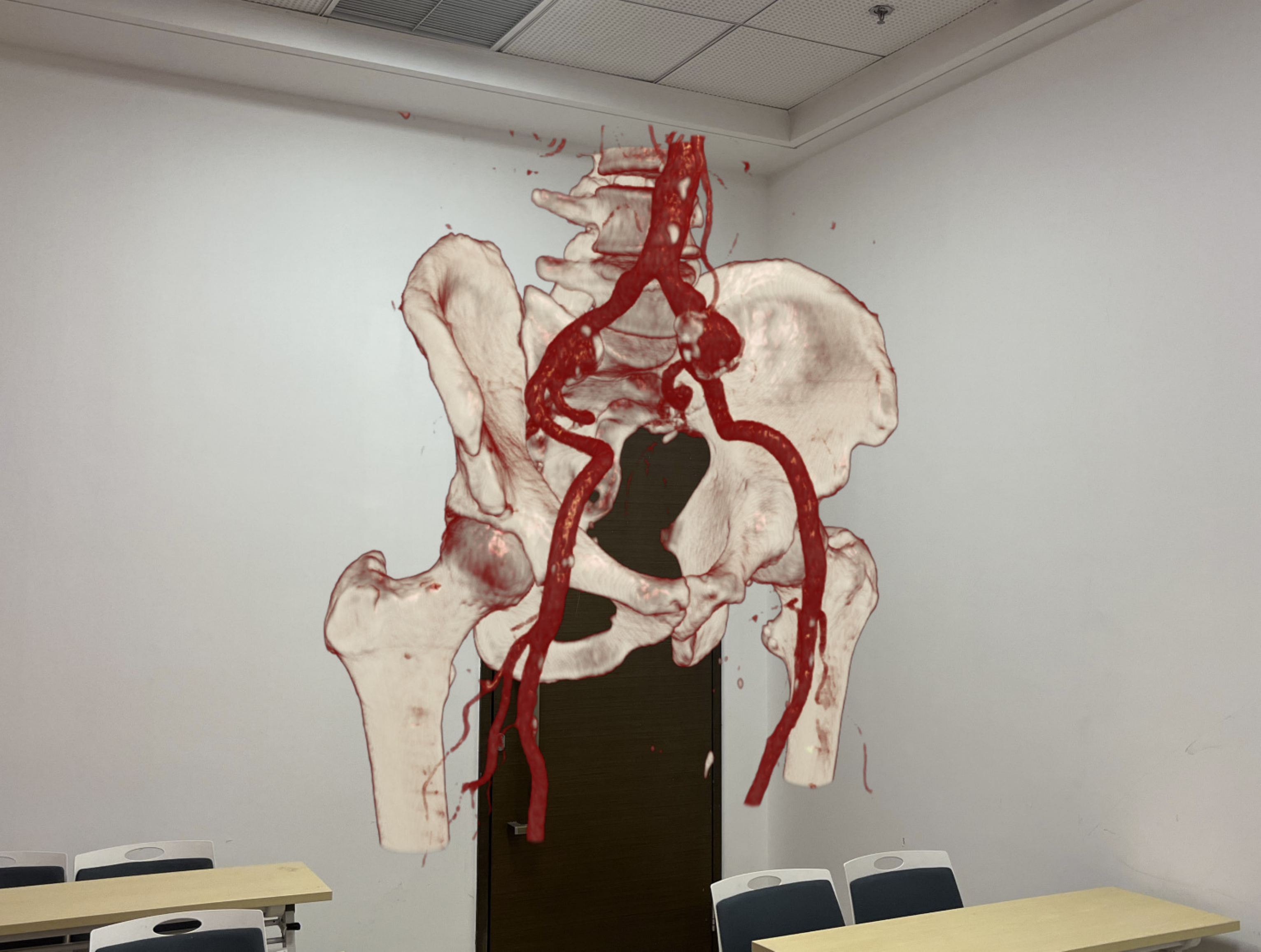}\hspace{-0.01in}
\includegraphics[width=0.312\linewidth]{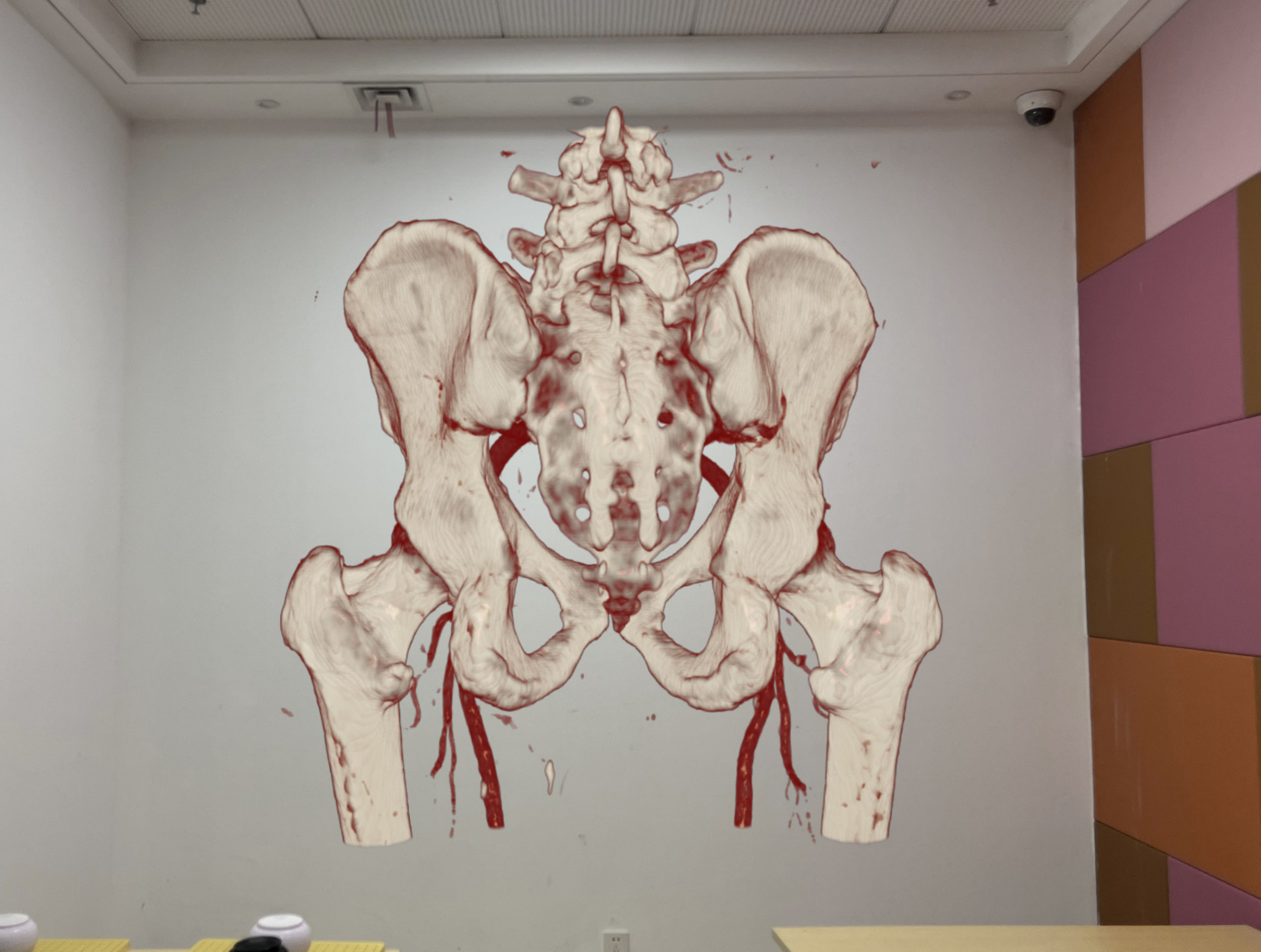}
\end{minipage}\\
\begin{minipage}[b]{0.95\linewidth}
\centering
\raisebox{0.28in}{\rotatebox{90}{\scalebox{.9}{\scriptsize{Ours}}}}
\includegraphics[width=0.312\linewidth]{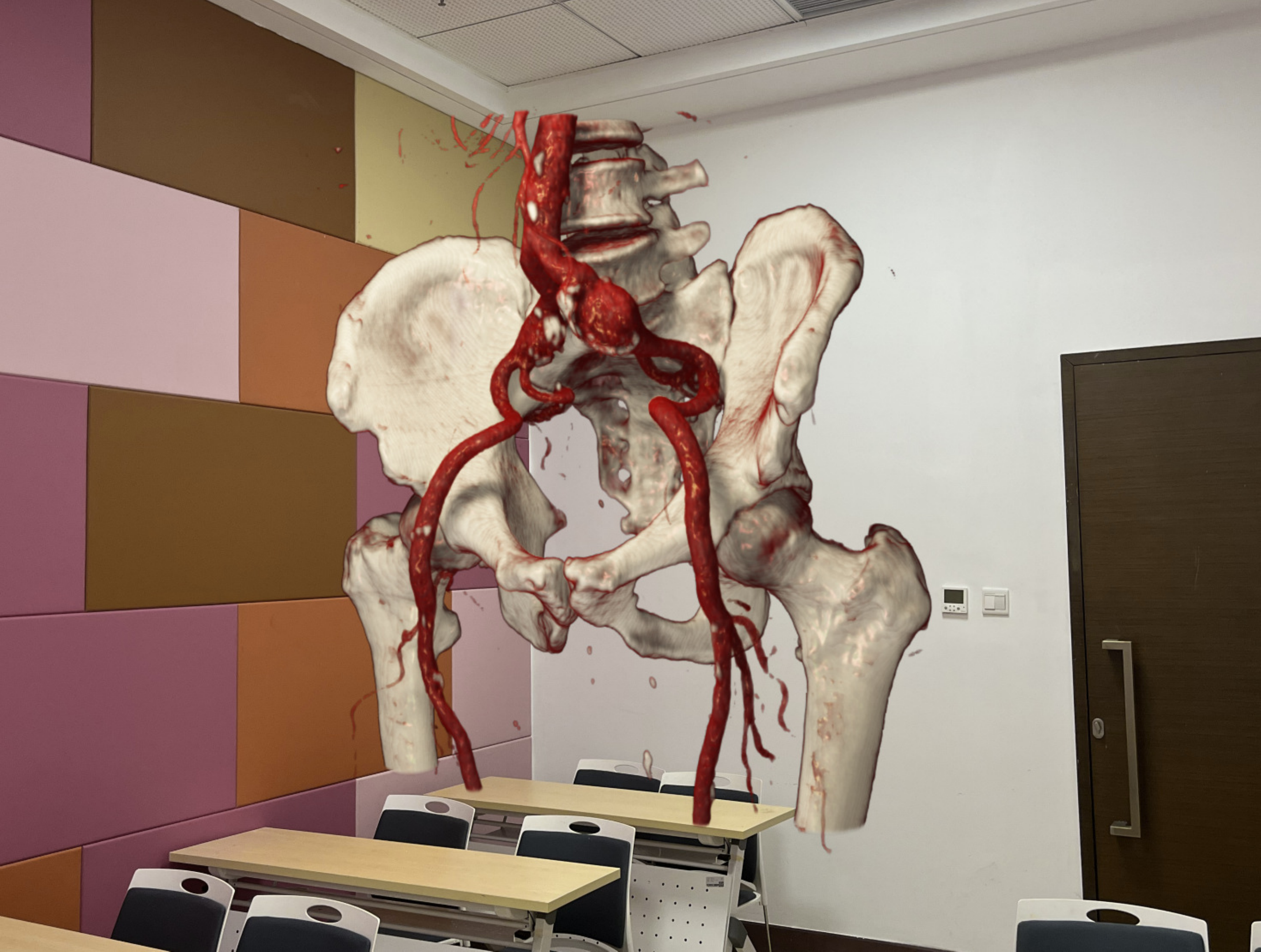}\hspace{-0.01in}
\includegraphics[width=0.312\linewidth]{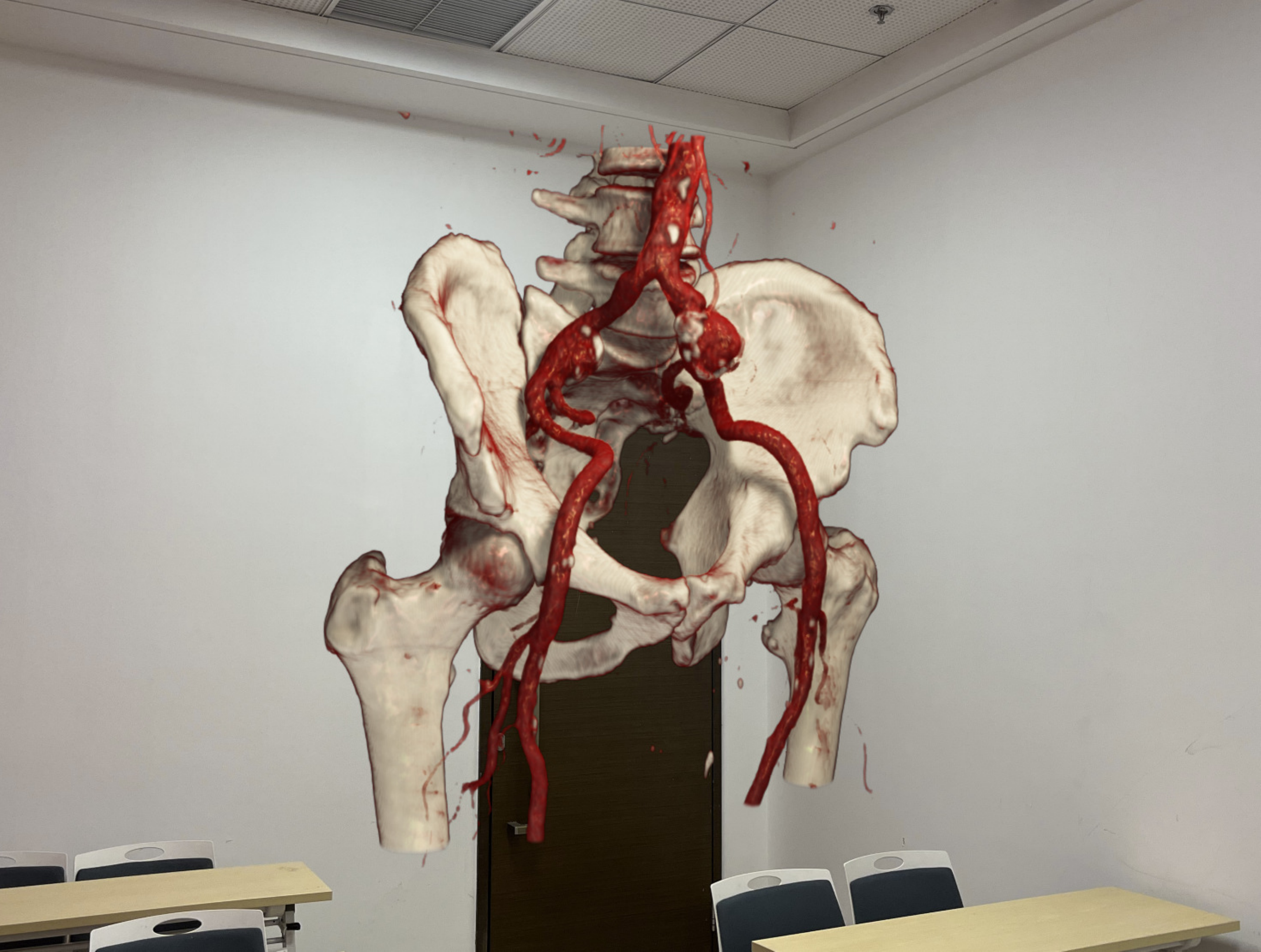}\hspace{-0.01in}
\includegraphics[width=0.312\linewidth]{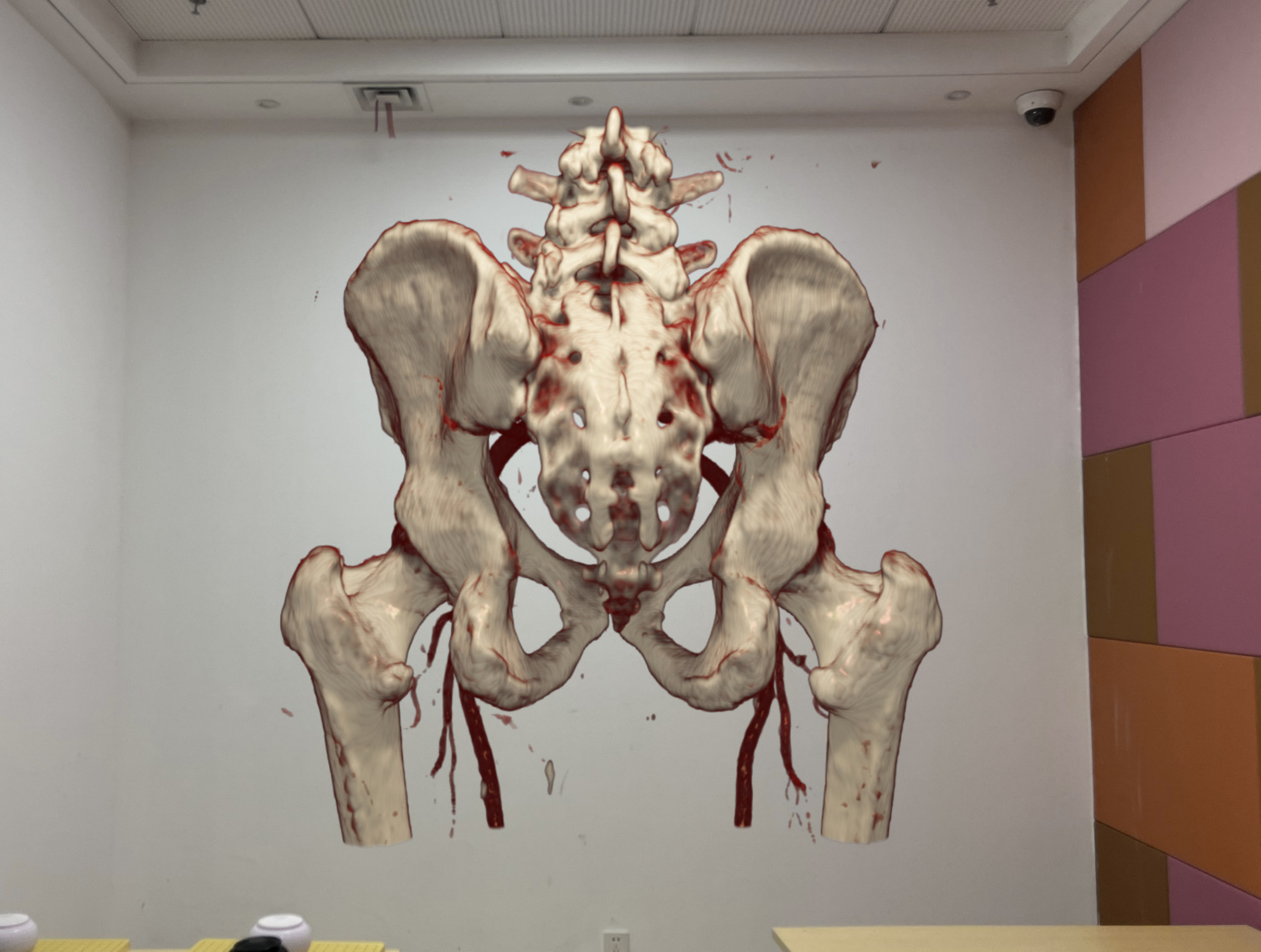}
\end{minipage}\\
\centerline{\scriptsize \quad View 0 \qquad\qquad\qquad View 1 \qquad\qquad\qquad View 2}
\caption{Visual comparison of MR rendering results generated from three different viewing angles using volumetric illumination models of our method and other existing approaches.}
\vspace{-0.20in}
\label{fig:temporalIMG}
\end{figure}

\textbf{Realistic DVR quality.} The ANOVA test of Q1 in Table \ref{tab:anova} showed significant differences between different techniques. The Bonferroni correction further showed that the proposed method ($M_{1} = 4.4$) performed significantly better than Li et al. \cite{Li20213d} ($M_{1} = 1.2$), Waschk et al. \cite{Waschk2020favr} ($M_{1} = 2.5$) and Rhee et al. \cite{Rhee2020} ($M_{1} = 3.6$). All participants agreed ($56\%$) or strongly agreed ($44\%$) that our method appeared realistic in MR. Nevertheless, high-fidelity DVR quality was inevitably accompanied by more rendering noise. As for Q2, since the methods of Li et al. \cite{Li20213d} and Waschk et al. \cite{Waschk2020favr} did not sample environment illumination, their responses had higher mean scores ($3.7$ and $3.5$ respectively) than our method ($M_{2} = 3.4$) and Rhee et al. \cite{Rhee2020} ($M_{2} = 3.2$). Although $p<0.05$, the Bonferroni correction further verified that no significant differences between testing techniques for Q2. Benefiting from the proposed spatio-temporal denoiser, our method produced an acceptable noise level without compromising realistic DVR.

\textbf{User perception of MR content.} The ANOVA test of this category (Q3 and Q4) in Table \ref{tab:anova} showed significant differences between different techniques. The Bonferroni correction further indicated that the proposed method ($M_{3} = 4.6$, $M_{4} = 4.2$) performed significantly better than Li et al. \cite{Li20213d} ($M_{3} = 1.2$, $M_{4} = 1.2$), Waschk et al. \cite{Waschk2020favr} ($M_{3} = 3.7$, $M_{4} = 1.4$) and Rhee et al. \cite{Rhee2020} ($M_{3} = 3.5$, $M_{4} = 3.8$). As for Q3, $80\%$ of the participants strongly disagreed that the method of Li et al. \cite{Li20213d} was adopted for observing the complex $3$D structures. Our method achieved higher Q3 scores than Waschk et al. \cite{Waschk2020favr} and Rhee et al. \cite{Rhee2020}, which indicated that the global shadows can better facilitate clear visualization of $3$D volumetric structures (Fig. \ref{fig:keyframes}). As for Q4, $76\%$ and $64\%$ of the participants strongly disagreed that the methods of Li et al. \cite{Li20213d} and Waschk et al. \cite{Waschk2020favr} matched this item. Several participants provided additional feedback that the MR content without spatially consistent shading effects was even more unnatural than observing the contents on a PC screen. Compare to our method (positivity rate: $92\%$), only $60\%$ of the participants agreed or strongly agreed that Rhee et al. \cite{Rhee2020} facilitated the harmonious integration of volumetric data with the surrounding real-world environment. We believe this is due to the unrealistic LDR illumination and the lack of necessary global shadows (Fig. \ref{fig:temporalIMG}). Compared to other categories, this category (Q3 and Q4) has larger $MS$-between values (Table. \ref{tab:anova}). It shows that our primary advantage in MR technology is the beneficial user perception of volumetric data in MR.

\textbf{MR interaction experience.} The ANOVA test of this category (Q5 and Q6) in Table \ref{tab:anova} showed no significant differences between different techniques. Since the testing techniques were not originally developed specifically for MR applications, all techniques used the same MR functional modules to make this assessment more objective and fair. Therefore, all testing techniques were robust without visible drift and jittering effects ($MS$-between = $0.09$ in Q5). Besides, most participants agreed that all testing techniques had indistinguishable interaction experiences ($MS$-between = $0.15$ in Q6). This was due to the limitation of the dual-screen refresh rate of HoloLens 2 ($60$ Hz). We also found that even if the DVR speed exceeded $60$ FPS, the actual MR display rate was hard to exceed $55$ FPS due to some in-built MR procedures of HoloLens 2 (e.g., hand-tracking and eye-tracking capabilities). From our observations, the MR display rate of all testing techniques dropped to $51\pm 2$ FPS. In the open-ended feedback, $5$ participants suggested that we develop more interaction tools (e.g., isosurface extraction and free cutting modules with the user's hands).

\section{Conclusion}
\label{sec:Conclusion}

A new MR framework for realistic DVR is proposed to improve user's sense of immersion during MR viewing. The proposed method first uses an HDR illumination estimation method to provide consistent real-world illumination for visualizing volumetric data. Based on the proposed spatio-temporal denoiser, the proposed DVR algorithm then utilizes the rendering results between adjacent frames and two screens of an MR HMD for denoising to improve DVR quality. To validate the effectiveness of our method, a comprehensive user study was conducted to evaluate the DVR quality of our method against existing volume illumination methods. Our experiment results demonstrate that the proposed method not only offers better volumetric data visualization quality in MR viewing but also exhibits satisfactory computational efficiency for real-time MR applications.

\bibliographystyle{abbrv-doi}

\vspace{0.20in}
\bibliography{References}

\end{document}